\documentclass[10pt,twocolumn,twoside]{IEEEtran}

\usepackage{epsf,times,amsmath,color}

\usepackage{epsfig,amsfonts,amsmath,graphicx}
\usepackage{subfigure,graphics,amssymb,amsxtra,color}
\usepackage{algorithm}
\usepackage{algorithmic}
\usepackage{flushend}

\usepackage[ansinew]{inputenc}

\usepackage{color,soul}
\usepackage{setspace}
\usepackage[hang,flushmargin,multiple]{footmisc}
\usepackage{epsfig}
\usepackage{wrapfig}
\usepackage{float}
\usepackage{amsfonts}
\usepackage{cite}
 \usepackage{acronym}
\usepackage{mymaths}

\newtheorem{lemma}{Lemma}
\newtheorem{theorem}{Theorem}

\renewcommand{\M}{\mathbf{\Phi}}

\renewcommand{\B}{\mathbf{B}}

\newcommand{\tr}{\mathrm{tr}}

\newcommand\rank{\mathrm{rank}}
\newcommand\MMSE{\text{MMSE}}
\newcommand\MSE{\text{MSE}}

\acrodef{CP}{cyclic prefix}
\acrodef{ZS}{zero pad suffix}
\acrodef{pdf}{probability density function}
\acrodef{iid}{independent and identically distributed}
\acrodef{BER}{bit error rate}
\acrodef{OFDM}{orthogonal frequency division multiplexing}
\acrodef{GSVD}{generalized singular value decomposition}
\acrodef{SVD}{singular value decomposition}
\acrodef{DMT}{discrete multitone}
\acrodef{ISI}{intersymbol interference}
\acrodef{ICI}{interchannel interference}
\acrodef{LOS}{line of sight}
\acrodef{NLOS}{non line of sight}
\acrodef{SNR}{signal-to-noise ratio}
\acrodef{PSNR}{peak signal-to-noise ratio}
\acrodef{SINR}{signal to interference plus noise ratio}
\acrodef{SIR}{signal to interference ratio}
\acrodef{MSE}{mean-squared error}
\acrodef{MIMO}{multiple-input multiple-output}
\acrodef{FFT}{fast Fourier transform}
\acrodef{IFFT}{inverse fast Fourier transform}
\acrodef{CDF}{cumulative distribution function}
\acrodef{CCDF}{complementary cumulative distribution function}
\acrodef{QAM}{quadrature amplitude modulation}
\acrodef{MMSE}{minimum mean-squared error}
\acrodef{LMMSE}{linear minimum mean-squared error}
\acrodef{SNR}{signal-to-noise ratio}
\acrodef{i.i.d.}{independent identically distributed}
\acrodef{SVD}{singular value decomposition}
\acrodef{MAP}{maximum \emph{a posteriori}}
\acrodef{MIMO}{multiple input multiple output}
\acrodef{OFDM}{orthogonal frequency division multiplexing}
\acrodef{CSI}{channel state information}
\acrodef{AWGN}{additive white Gaussian noise}
\acrodef{CDF}{cumulative distribution function}
\acrodef{KKT}{Karush-Kuhn-Tucker}
\acrodef{PDP}{power delay profile}
\acrodef{QPSK}{quadrature phase-shift keying}
\acrodef{CS}{compressive sensing}
\acrodef{GMM}{Gaussian mixture model}
\acrodef{OMP}{orthogonal matching pursuit}
\acrodef{EM}{expectation maximization}

\title{Reconstruction of Signals Drawn from a Gaussian Mixture via Noisy Compressive Measurements}

\author{\normalsize Francesco~Renna,~\IEEEmembership{Member,~IEEE,}
        Robert~Calderbank,~\IEEEmembership{Fellow,~IEEE,}
        Lawrence~Carin,~\IEEEmembership{Fellow,~IEEE,}
        and~Miguel~R.~D.~Rodrigues,~\IEEEmembership{Member,~IEEE}
\thanks{This paper was presented in part at the 2013 IEEE Global Conference on Signal and Information Processing. }
\thanks{The work of F. Renna was supported by Funda\c{c}\~{a}o para a Ci\^encia e a Tecnologia through the research project CMU-PT/SIA/0026/2009. The work of M. R. D. Rodrigues was supported by the EPSRC through the research grant EP/K503459/1. The work of R. Calderbank and L. Carin was partially supported by the Defense Advanced Research Projects Agency (DARPA) under the KeCom program, and also by the Office of Naval Research (ONR) and by the Air Force Office of Scientific Research under the Complex Networks Program. This work was also supported by the Royal Society International Exchanges Scheme IE120996.}
\thanks{F. Renna is with the Instituto de Telecomunica\c{c}\~{o}es and the Departamento de Ci\^{encia} de Computadores, Faculdade de Ci\^{e}ncias da Universidade do Porto, Porto, Portugal (e-mail: frarenna@dcc.fc.up.pt).}
\thanks{R. Calderbank and L. Carin are with the Department of Electrical and Computer Engineering, Duke University, Durham NC, USA (e-mail: \{robert.calderbank, lcarin\}@duke.edu).}
\thanks{M. R. D. Rodrigues is with the Department of E\&EE, University College London, London, UK (email: m.rodrigues@ucl.ac.uk).}
}

\begin{document}

\maketitle

\begin{abstract}
This paper determines to within a single measurement the minimum number of measurements required to successfully reconstruct a signal drawn from a Gaussian mixture model in the low-noise regime. The method is to develop upper and lower bounds that are a function of the maximum dimension of the linear subspaces spanned by the Gaussian mixture components. The method not only reveals the existence or absence of a minimum mean-squared error (MMSE) error floor (phase transition) but also provides insight into the MMSE decay via multivariate generalizations of the MMSE dimension and the MMSE power offset, which are a function of the interaction between the geometrical properties of the kernel and the Gaussian mixture. These results apply not only to standard linear random Gaussian measurements but also to linear kernels that minimize the MMSE. It is shown that optimal kernels do not change the number of measurements associated with the MMSE phase transition, rather they affect the sensed power required to achieve a target MMSE in the low-noise regime. Overall, our bounds are tighter and sharper than standard bounds on the minimum number of measurements needed to recover sparse signals associated with a union of subspaces model, as they are not asymptotic in the signal dimension or signal sparsity. 

\end{abstract}


\section{Introduction}
\label{introduction}


The foundation of the digital revolution is the Shannon-Nyquist theorem, which provides a theoretical basis for digital processing of analog signals: it states that the sampling rate should be at least twice the Fourier bandwidth of the signal. The discrete-time representation of a continuous-time signal, which lies at the heart of analogue to digital conversion, offers the means to resilient data communication, storage and processing.

It has been recognized recently that the so-called Nyquist rate can be excessive in various emerging applications~\cite{Healy08mag,Lustig08mag,Duarte08mag}: this -- in addition to representing a burden to analog-to-digital converters~\cite{ADconv09,Dataconv09} -- can also lead to a huge number of samples that compromise communications, storage and processing resources. Modern acquisition systems thus adopt a two-step approach that involves both an analog-to-digital conversion operation, whose purpose is to convert the information-bearing signal from the analogue to the digital domain, and a compression operation whose purpose is to offer succinct (near lossless) representations of the data.

There has been a recent emergence of a new sensing modality, emblematically known as Compressive Sensing (CS)~\cite{Candes06,CandesTao06,Donoho06}, that offers the means to simultaneously sense and compress a signal without any loss of information (under appropriate conditions on the signal model and measurement process). The sensing process is based on the projection of the signal of interest onto a set of vectors, which can be either constituted randomly~\cite{Candes06,CandesTao06,Donoho06,Candes05,Baraniuk08} or designed~\cite{Ashok08,Baheti09}, and the recovery process is based on the resolution of an inverse problem. It is well known that the reconstruction of an $n$-dimensional signal that admits an $s$-sparse representation in some orthonormal basis or frame, via $\ell_0$-pseudonorm minimization algorithms, requires only $s+1$ noiseless measurements~\cite{Venkataramani98,Baron06}. However, there are no tractable algorithms able to solve such a minimization problem. On the other hand, $\ell_1$ minimization methods~\cite{Candes06IT} or iterative methods, like greedy matching pursuit~\cite{Mallat93,Chen98,Tropp10}, provide reliable reconstruction with overwhelming probability  with only $\mathcal{O}(s \log(n/s))$ linear random measurements or projections ~\cite{Candes06,Donoho06,Baraniuk08}.


However, even in early CS studies, it has been recognized that it is possible to derive better compression performance, in terms of the minimum number of measurements necessary to achieve perfect or nearly perfect reconstruction, by leveraging the fact that signals often obey models with additional structure beyond conventional sparsity. Some popular models that capture such additional structure include the union of subspaces~\cite{Blumensath09,Stojnic09,Eldar09,Eldar10}, wavelet trees~\cite{Blumensath09,Baraniuk10} or manifolds~\cite{Baraniuk09,Chen10}. Within the union of subspaces model, the source signal is assumed to belong to one out of a collection of $K$ subspaces with dimension less than or equal to $s$. 
Recovery of a signal in a union of subspaces is equivalent to the reconstrution of a block-sparse signal, when the individual subspaces in the union of subspaces model are decomposable as the direct sum of a given number of lower dimensional subspaces~\cite{Eldar09}. The minimum number of measurements required for reliable reconstruction in such scenarios has been shown to be of the order $\mathcal{O}(s + \log(2 K))$~\cite{Blumensath09} when using mixed $\ell_2/\ell_1$-norm minimization~\cite{Eldar09}. 
On the other hand, tree models, where the non-zero coefficients of the source signal are known to be gathered into a rooted, connected, tree structure, can describe the most relevant wavelet coefficients of piecewise smooth signals or images~\cite{Crouse98}. In this case, the number of measurements required for reliable reconstruction is of order $\mathcal{O}(s)$~\cite{Baraniuk10} by using a model-based version of the CoSaMP algorithm~\cite{Needell09}. Finally, in the case of Riemannian manifolds, the minimum number of random projection measurements needed for reliable reconstruction has been derived to be of the order $\mathcal{O}(s \log(n V R \tau^{-1}))$, where $s$ is the dimension of the manifold, $\tau^{-1}$ being its condition number, $V$ being its volume and $R$ being its geodesic covering regularity~\cite{Baraniuk09}.


Another very useful structured model is the \ac{GMM}~\cite{Chen10,Yu11,YuSapiro12,DuarteC13}. GMMs are typically used in conjunction with the Bayesian CS formalism that entails the use of statistical descriptions of the source and a statistical description of the measurement system in order to perform reconstruction~\cite{Ji08}. One important feature of these models relates to the existence of efficient and optimal inversion procedures, which can be expressed analytically in closed form~\cite{Chen10}. The other important feature -- in addition to the approximation of any distribution with arbitrary precision ~\cite{Sorenson71} -- relates to the fact that these models have also been shown to provide state-of-the-art results in various practical problems~\cite{YuSapiro12}. These include problems in image processing such as interpolation, zooming, deblurring~\cite{Yu11,YuSapiro12,DuarteC13} and dictionary learning~\cite{Chen10}.

A \ac{GMM} also relates to the other well-known structured models in the literature~\cite{Blumensath09,Stojnic09,Eldar09,Eldar10,Baraniuk09,Chen10}. For example, the \ac{GMM} can be seen as a Bayesian counterpart of the union of subspaces model (assuming each GMM mixture component has a near-low-rank covariance matrix). In fact, a signal drawn from a \ac{GMM} lies in a union of subspaces, where each subspace corresponds to the image of each class-conditioned covariance matrix in the model\footnote{More generally, a signal drawn from a \ac{GMM} model lies in a union of affine spaces rather than linear subspaces, where each affine space is associated with the mean and covariance of each class in the \ac{GMM} model.}. In addition, a low-rank \ac{GMM} can also be seen as an approximation to a compact manifold. Compact manifolds can be covered by a finite collection of topological disks that can be represented by high-probability ellipsoids living on the principal hyperplanes corresponding to the different components of a low-rank \ac{GMM}~\cite{Chen10}. 
However, we emphasize that adopting a \ac{GMM} \emph{in lieu} of the other structured models has a specific advantage. Reconstruction of a signal drawn from a \ac{GMM} from compressive linear measurements in Gaussian noise can be very effectively performed via a closed-form inversion formula~\cite{Chen10}.

As such, and also in view of its practical relevance, this paper studies in detail the behavior of the \ac{MMSE} associated with the reconstruction of a signal drawn from a GMM, based on a set of linear and noisy compressive measurements. We consider the asymptotic regime of low-noise, which is relevant in various signal and image processing scenarios~\cite{Carson12,Chen12}. The emphasis is to understand, as a function of the properties of the linear measurement kernel and the Gaussian mixture, whether the \ac{MMSE} converges or does not converge to zero as the noise power converges to zero, i.e. the \ac{MMSE} phase transition. The main contributions are:

\begin{itemize}

\item A bound on the number of linear random measurements that are necessary to reconstruct perfectly a signal drawn from a \ac{GMM} in the low-noise regime;

\item A bound on the number of linear random measurements that are sufficient to reconstruct perfectly a signal drawn from a GMM in the low-noise regime, by analyzing the MMSE performance of a (sub-optimal) classification and reconstruction strategy;

\item Generalization of the bounds on the number of measurements that are necessary and/or sufficient to reconstruct a signal drawn from a GMM based on a set of noisy compressive measurements, considering the scenario where the linear measurement kernel is constituted randomly and then extended to the case for which the linear measurement kernel is \emph{designed} to minimize the mean-squared error;

\item Characterization, whenever possible, of a more refined behavior of the low-noise asymptotics of the MMSE, that portray the existence or absence of an MMSE error floor (the phase transition) as well as the MMSE decay, as a function of the geometry of the kernel and the geometry of the Gaussian mixture.

\end{itemize}

Overall, this contribution offers an analysis of the reconstruction performance and associated phase transitions that, and in contrast with other results in the CS literature (e.g.~\cite{Candes05,Candes06,Candes07,Donoho09,Donoho11,Blumensath09,Eldar09,Eldar10,Baraniuk10,Baraniuk09,Wu12,Reeves12,Tulino12}), is non-asymptotic in the signal dimension or the signal sparsity. Recent works have also proposed the use of message passing and belief propagation methods to increase the speed of reconstruction algorithms~\cite{Sarvotham06,Guo06,Donoho09pnas,Bayati11,Rangan11}. However, these approaches have also been studied under the large-system assumption. In particular, message passing methods are proved to be computationally efficient in large-scale applications, while guaranteeing reliable reconstruction with a number of measurements $n$ the same order as $\ell_1$-norm minimization~\cite{Sarvotham06,Donoho09pnas}. In fact, message passing algorithms are shown to be equivalent to \ac{MMSE} estimation in the asymptotic large-system limit, when the projection matrix is sparse and the inputs are \ac{i.i.d.} with arbitrary distribution~\cite{Guo06}. 

On the other hand, the analysis proposed in this paper is based on models that naturally incorporate memory rather than memoryless models, as opposed to previous contributions in the literature on the information-theoretic characterization of {CS}~\cite{Wu12,Reeves12,Tulino12}. Memoryless models, however, have been characterized in terms of the reconstruction {MMSE} in the large-system limit. In particular, \cite{Guo09All} shows that, in the large-system limit, the overall \ac{MSE} can be decoupled into the \ac{MSE} relative to the reconstruction of the individual elements of the sparse signal, and such value admits a single-letter characterization. This analysis is justified on the basis of a heuristic method from statistical physics, the \emph{replica method}, but it can be proved rigorously for the case of sparse measurement matrices. The replica method has been shown to provide bounds that are in agreement with the exact analysis for the case of sparsity-pattern recovery in the low-noise regime~\cite{Reeves12CISS}, and it has been used also to evaluate the \ac{MSE} associated to the \ac{MAP} estimator in the large system limit~\cite{Rangan11arx}.


The remainder of the paper is organized as follows: Section~\ref{par:sysmodel} introduces the system model and the main performance quantities and definitions. The impact of a random linear measurement kernel on the behavior of the MMSE and its phase transition is investigated first for a signal drawn from a Gaussian distribution in Section~\ref{par:Gaussinputs}, to develop essential intuition; we then consider a signal drawn from a mixture of Gaussian distributions in Section~\ref{par:GMMinputs}. The impact on the phase transition of designing the measurement kernel is then illustrated in Section~\ref{sec:Design}. Section~\ref{par:numres} exhibits various numerical results both with synthetic and real data, illustrating the main operational features of the problem. Section~\ref{par:conclusion} summarizes the main contributions. For completeness, Appendix~\ref{app:lemmas} collects a number of useful Lemmas that are relevant for the proofs of the main results reported in Appendices \ref{app:Gauss} and \ref{app:UpGMM}.

We use the following notation: boldface upper-case letters denote matrices (${\bf X}$) and boldface lower-case letters denote column vectors (${\bf x}$); the context defines whether the quantities are deterministic or random. The symbols ${\bf I}_n$ and $\mathbf{0}_{m \times n}$ represent the identity matrix of dimension $n \times n$ and the all-zero-entries matrix of dimension $m \times n$, respectively (subscripts will be dropped whenever the dimensions are clear from the context). The expressions $\left(\cdot\right)^{\dag}$, $\tr(\cdot)$,  $\rank(\cdot)$ represent the transpose, trace and the rank operators, respectively. $\mathrm{Im}(\cdot)$ and $\mathrm{Null}(\cdot)$ denote the (column) image and null space of a matrix, respectively, $(\cdot)^\perp$ denotes the orthogonal complement of a linear subspace, and $\dim (\cdot )$ denotes the dimension of a linear subspace. $\E{\cdot}$ represents the expectation operator. The Gaussian distribution with mean $\boldsymbol{\mu}$ and covariance matrix $\mathbf{\Sigma}$ is denoted by $\mathcal{N}(\boldsymbol{\mu},\mathbf{\Sigma})$.

\section{System model}
\label{par:sysmodel}

We study the problem associated with the reconstruction of a source signal $\mathbf{x} \in \mathbb{R}^n$ from a set of $\ell < n$ noisy, linear projections $\mathbf{y} \in \mathbb{R}^{\ell}$ where
\begin{equation}
\mathbf{y} =   \mathbf{\Phi} \, \mathbf{x} + \mathbf{w},
\label{eq:model}
\end{equation} 
and $\mathbf{\Phi} \in \mathbb{R}^{\ell \times n}$ is the linear measurement kernel\footnote{Throughout the paper, we will refer to $\mathbf{\Phi}$ as the sensing matrix, measurement matrix and kernel, interchangeably.} and $\mathbf{w} \sim \mathcal{N}(\mathbf{0}, \sigma^2 \cdot \mathbf{I}_\ell )$ is zero-mean, \ac{AWGN}. We consider in Sections \ref{par:Gaussinputs} and \ref{par:GMMinputs} random measurement kernel designs, where the entries of $\mathbf{\Phi}$ are drawn \ac{i.i.d.} from a zero-mean, fixed-variance, Gaussian distribution, which is common in the CS literature~\cite{Candes06,Donoho06}. However, we also consider in Section~\ref{sec:Design} design of the measurement kernel $\mathbf{\Phi}$, that aims to minimize the reconstruction error.

In this work we also concentrate on two particular distributions for the source vector: a Gaussian distribution and a \ac{GMM} distribution (of course, the former is a special case of the latter). For a Gaussian source vector,  $\mathbf{x} \sim \mathcal{N}( \boldsymbol{\mu}_{\mathbf{x}},\mathbf{\Sigma}_{\mathbf{x}})$ where $\boldsymbol{\mu}_{\mathbf{x}}$ represents the mean and $\mathbf{\Sigma}_{\mathbf{x}}$ represents the (possibly rank deficient) covariance matrix.  We denote the eigenvalue decomposition of the positive semidefinite covariance matrix
\begin{IEEEeqnarray}{rCl}
\mathbf{\Sigma}_{\mathbf{x}} & = &  \mathbf{U}_{\mathbf{x}} \mathbf{\Lambda}_{\mathbf{x}} \mathbf{U}_{\mathbf{x}}^\dag =  \mathbf{U}_{\mathbf{x}} \,  \mathrm{diag}(\lambda_{x_1}, \ldots, \lambda_{x_s},0,\dots, 0)  \,   \mathbf{U}_{\mathbf{x}}^\dag, \IEEEeqnarraynumspace
\end{IEEEeqnarray}
where the orthogonal matrix $\mathbf{U}_{\mathbf{x}}$ contains the eigenvectors of $\mathbf{\Sigma}_{\mathbf{x}}$, the diagonal matrix $\mathbf{\Lambda}_{\mathbf{x}} = \mathrm{diag}(\lambda_{x_1}, \ldots, \lambda_{x_s},0,\dots, 0)$ contains the eigenvalues of $\mathbf{\Sigma}_{\mathbf{x}}$, $\lambda_{\mathbf{x}_1} \geq \ldots \geq \lambda_{\mathbf{x}_s} > 0$, and $s= \rank(\mathbf{\Sigma}_{\mathbf{x}})$ represents the rank of $\mathbf{\Sigma}_{\mathbf{x}}$. 



For a \ac{GMM} source, $\mathbf{x} \sim \sum_k p_k \, \mathcal{N}( \boldsymbol{\mu}_{\mathbf{x}}^{(k)},\mathbf{\Sigma}_{\mathbf{x}}^{(k)})$, so that the source vector is assumed to be drawn from one out of $K$ different classes with probability $p_k$, $k=1,\ldots,K$ where the distribution of the source vector conditioned on the class $k$ is Gaussian with mean $\boldsymbol{\mu}_{\mathbf{x}}^{(k)}$ and (possibly rank deficient) covariance matrix $\mathbf{\Sigma}_{\mathbf{x}}^{(k)}$. We let $s_k = \rank (\mathbf{\Sigma}_{\mathbf{x}}^{(k)})$, $s_{km}=\rank (\mathbf{\Sigma}_{\mathbf{x}}^{(k)} + \mathbf{\Sigma}_{\mathbf{x}}^{(m)} )$ and $s\sub{max} = \max_k s_k$. 

A low-rank modeling approach, where $s_k < n$ for some or all $k$, is the basis of the theory developed in Sections~\ref{par:Gaussinputs}, \ref{par:GMMinputs} and \ref{sec:Design}. This implies that a realization of the source signal lies on one out of the $K$ affine subspaces associated with the translation by the mean vector $\boldsymbol{\mu}_{\mathbf{x}}^{(k)}$ of the subspaces corresponding to the images of the class-conditioned covariance matrices $\mathbf{\Sigma}_{\mathbf{x}}^{(k)}$. Therefore, a signal drawn from a \ac{GMM} model can also be seen to lie on a union of subspaces (or affine subspaces)~\cite{Eldar09}. However, the fact that natural signals and images are not always exactly low-rank but rather ``approximately'' low-rank is also discussed in the sequel (see Section~\ref{par:RealData}).



We use the \ac{MMSE} to assess the level of distortion incurred in the reconstruction of the original source vector ${\bf x}$ from the projections vector ${\bf y}$ in the CS model in (\ref{eq:model}), which is given by:
\begin{equation}
\MMSE =  \E{\|\mathbf{x} - \hat{\mathbf{x}}(\mathbf{y})\|^2},
\end{equation}
where
\begin{equation}
\hat{\mathbf{x}}(\mathbf{y}) = \E{\mathbf{x}|\mathbf{y}}.
\end{equation}


We focus on the characterization of the behavior of the \ac{MMSE} in the low-noise regime, i.e., for $\sigma^2 \to 0$, which represents the regime with most operational relevance in many signal and image processing applications (the noise magnitude typically considered for these applications is $\sigma^2 = -60$\, dB~\cite{Carson12,Chen12}). This includes the characterization of an asymptotic expansion of the \ac{MMSE} as $\sigma^2 \to 0$ together with the characterization of the phase transition of the \ac{MMSE} as $\sigma^2 \to 0$. The \ac{MMSE} phase transition, in line with other results in the literature~\cite{Wu12,Donoho11}, corresponds to the minimum number of measurements that guarantee perfect reconstruction in the low-noise regime.

\section{Gaussian sources}
\label{par:Gaussinputs}

We first consider the characterization of the \ac{MMSE} phase transition and a low-noise \ac{MMSE} expansion associated with a linearly and compressively measured signal drawn from a Gaussian source. Such characterizations, which can be crisply expressed in terms of the geometry of the measurement kernel, the geometry of the source and their interplay, pave the way to the characterization of the \ac{MMSE} phase transition associated with \ac{GMM} sources.

For a Gaussian source with mean $\boldsymbol{\mu}_{\mathbf{x}}$ and covariance matrix $\mathbf{\Sigma}_{\mathbf{x}}$, in the presence of zero-mean, additive Gaussian noise with covariance matrix $\sigma^2 \mathbf{I}$, the conditional mean estimator can be expressed as follows~\cite{Hassibi}:
\begin{equation}
\hat{\mathbf{x}}(\mathbf{y}) = \mathcal{W}(\mathbf{y})  = \boldsymbol{\mu}_x +  \mathbf{W}(\mathbf{y}  - \mathbf{\Phi} \boldsymbol{\mu}_x),
\label{WGauss}
\end{equation}
where $\mathbf{W} =  \mathbf{\Sigma}_{\mathbf{x}} \mathbf{\Phi}^\dag (\sigma^2 \mathbf{I} + \mathbf{\Phi} \mathbf{\Sigma}_{\mathbf{x}}\mathbf{\Phi}^\dag)^{-1}$ corresponds to the Wiener filter associated with a Gaussian source with mean zero and covariance $\mathbf{\Sigma}_{\mathbf{x}}$, and the \ac{MMSE} can also be expressed in closed form as follows~\cite{Hassibi}:
\begin{equation}
\MMSE^{\sf G}(\sigma^2) = \tr \left(  \mathbf{\Sigma}_{\mathbf{x}}  -  \mathbf{\Sigma}_{\mathbf{x}} \mathbf{\Phi}^\dag \left( \sigma^2 \mathbf{I} +  \mathbf{\Phi} \mathbf{\Sigma}_{\mathbf{x}} \mathbf{\Phi}^\dag   \right)^{-1} \mathbf{\Phi} \mathbf{\Sigma}_{\mathbf{x}}   \right),
\label{eq:MMSEGauss}
\end{equation}
where we explicitly highlight the dependence of the MMSE on the measurement noise variance $\sigma^2$, assuming a fixed Gaussian source with covariance $\mathbf{\Sigma}_\mathbf{x}$. 
The following theorem now reveals the asymptotic behavior of the \ac{MMSE} in the low-noise regime. We define
\begin{equation}
\mathbf{\Sigma} = \mathbf{\Sigma}_{\mathbf{x}}^{\frac{1}{2}} \mathbf{\Phi}^\dag \mathbf{\Phi} \mathbf{\Sigma}_{\mathbf{x}}^\frac{1}{2} = \mathbf{U} \mathbf{\Lambda} \mathbf{U}^{\dag},
\label{eq:sigma}
\end{equation}
where $\mathbf{\Sigma}_{\mathbf{x}}^{\frac{1}{2}}$ is the (positive semidefinite) matrix square root of $\mathbf{\Sigma}_{\mathbf{x}}$, $\mathbf{U}$ is an orthogonal matrix that contains the eigenvectors of $\mathbf{\Sigma}$, $\mathbf{\Lambda} = \mathrm{diag}(\sqrt{\lambda_{1}}, \ldots, \sqrt{\lambda_{\ell'}},0,\dots, 0)$ is a diagonal matrix that contains the eigenvalues of $\mathbf{\Sigma}$, $\lambda_1 \geq \ldots \geq \lambda_{\ell'} > 0$ and $\ell' = \rank(\mathbf{\Sigma})$. Note also that, when the entries of the measurement kernel $\mathbf{\Phi}$ are drawn \ac{i.i.d.} from a zero-mean, fixed-variance, Gaussian distribution, the rank of $\mathbf{\Sigma}$ is equal to $\ell' = \rank(\mathbf{\Sigma}) = \min \{  s,\ell  \} $, with probability 1.
\begin{theorem}
\label{theo:Gauss}
Consider the linear measurement model in (\ref{eq:model}) where $\mathbf{x} \sim \mathcal{N}(\boldsymbol{\mu}_{\mathbf{x}}, \mathbf{\Sigma}_{\mathbf{x}})$, $s= \rank (\mathbf{\Sigma}_\mathbf{x})$ and $\ell$ is the number of measurements. 
The first-order, low-noise expansion of the \ac{MMSE} is given by:
\begin{equation}
\MMSE^{\sf G}(\sigma^2) = \mathcal{M}^{\sf G}_{\infty} + {\mathcal{D}^{\sf G }} \cdot \sigma^2 + { o}\left( \sigma^2 \right).
\label{eq:expGauss}
\end{equation}
The zero-order term in the expansion, which relates to the \ac{MMSE} floor, is given by:
\begin{equation}
 \mathcal{M}_{\infty}^{\sf G} = \lim_{\sigma^2 \rightarrow 0} \MMSE^{\sf G}(\sigma^2) = \sum_{i=\ell' +1}^s  {\mathbf{u}}_i^\dag \mathbf{\Sigma}_{\mathbf{x}} {\mathbf{u}}_i ,
\label{eq:GaussFloor}
\end{equation}
where the vectors ${\mathbf{u}}_{\ell'+1},\ldots,{\mathbf{u}}_{s}$ form an orthonormal basis of the linear subspace $\mathrm{Null}(\mathbf{\Sigma})  \cap \mathrm{Null}(\mathbf{\Sigma}_{\mathbf{x}})^\perp$, and the coefficient of the first-order term in the expansion is given by:
\begin{equation}
 \mathcal{D}^{\sf G} = \sum_{i=1}^{\ell'} \frac{1}{\lambda_i} \mathbf{u}_i^{\dag} \mathbf{\Sigma}_{\mathbf{x}} \mathbf{u}_i,
 \label{eq:GaussD}
 \end{equation}
where $\mathbf{u}_1,\ldots, \mathbf{u}_{\ell'}$ are eigenvectors of $\mathbf{\Sigma}$ corresponding to the positive eigenvalues $\lambda_1,\ldots, \lambda_{\ell'}$.
\end{theorem}

\begin{IEEEproof}
The proof of this Theorem is provided in Appendix~{\ref{app:Gauss}}.
\end{IEEEproof}
The value of the zero-order term is clearly zero if $\ell \geq s$ and it is non-zero otherwise with its value dictacted by the interaction of the geometry of the source with the geometry of the measurement kernel, as portrayed via (\ref{eq:GaussFloor}) and (\ref{eq:sigma}). The value of the coefficient of the first-order term is non-zero with its value depending as well on the interaction of the description of the source and of the kernel.

Theorem \ref{theo:Gauss} and (\ref{eq:expGauss}), (\ref{eq:GaussFloor}) and (\ref{eq:GaussD}) then lead to the conclusions:

\begin{itemize}

\item  When $\ell < s$ the number of measurements is not sufficient to capture the full range of the source, so that the reconstruction is not perfect ($\mathcal{M}^{\sf G}_{\infty} \neq 0$). On the other hand, when $\ell \geq s$ such a number of measurements capture completely the source information leading to perfect reconstruction ($\mathcal{M}^{\sf G}_{\infty} = 0$) in the low-noise regime. That is, one requires the number of linear random measurements to be greater than or equal to the dimension of the subspace spanned by the source for the {phase transition} to occur. 



\item When $\ell \geq s$, the rate of decay of the \ac{MMSE} is $\mathcal{O} \left(\sigma^2\right)$ as $\sigma^2 \to 0$, as in the scalar case~\cite{Wu11}. 
On the other hand, the power offset of the \ac{MMSE} on a $10 \cdot \log_{10} \frac{1}{\sigma^2}$--scale  is dictated by the quantity $10 \cdot \log_{10} \mathcal{D}^{\sf G}$. In fact, the quantity $\mathcal{D}^{\sf G}$, which represents the multivariate Gaussian counterpart of the \ac{MMSE} dimension put forth in~\cite{Wu11}, distinguishes \ac{MMSE} expansions associated with different realizations of the measurement kernel and different source covariances.

\item Of particular interest, the presence or absence of a \ac{MMSE} floor depends only on the relation between the number of measurements $\ell$ and the rank of the source covariance $s$. On the other hand, the exact value of the \ac{MMSE} floor (when $\ell < s$) and the \ac{MMSE} power offset (when $\ell \geq s$) depends on the relation between the geometry of the random measurement kernel and the geometry of the source.

\end{itemize}


\section{GMM sources}
\label{par:GMMinputs}

We are now ready to consider the characterization of \ac{MMSE} phase transitions associated with a linearly and compressively measured signal drawn from a GMM source. In particular, and in view of the lack of a closed-form tractable  \ac{MMSE} expression, we derive necessary and sufficient conditions for the phase transitions to occur via bounds to the \ac{MMSE}  for \ac{GMM} sources, which will be denoted by the symbol $\MMSE^{\sf GM}(\sigma^2)$.



\subsection{Necessary condition}
\label{par:lowerGMM}


The necessary condition on the number of random linear measurements (components of $\mathbf{\Phi}$ constituted at random) for the \ac{MMSE} phase transition to occur is based on the analysis of a lower bound to the \ac{MMSE}. 
We express the lower bound in terms of the Gaussian \ac{MMSE} associated to the actual class from which each signal realization is drawn. Namely, we have 
\begin{IEEEeqnarray}{rCl}
\MMSE^{\sf GM} (\sigma^2)&=& \E{\|  \mathbf{x} -   \hat{\mathbf{x}}(\mathbf{y}) \|^2 } \\
\label{equal1}
 &=&\sum_{k=1}^K p_k \, \E{\|  \mathbf{x} -   \hat{\mathbf{x}}(\mathbf{y}) \|^2 | c=k} \\
 \label{inequal2}
 &\geq & \sum_k p_k \, \MMSE^{\sf G}_k(\sigma^2)  =  \MSE_{\sf LB}(\sigma^2) ,\IEEEeqnarraynumspace
\end{IEEEeqnarray}
where $\MMSE^{\sf G}_k(\sigma^2) $ denotes the \ac{MMSE} associated with the reconstruction of Gaussian signals ${\bf x}$ in class $c=k$ from the measurement vector ${\bf y}$. Note that the equality in (\ref{equal1}) is due to the total probability formula and the inequality in (\ref{inequal2}) follows from the optimality of the \ac{MMSE} estimator (\ref{WGauss}) for a single Gaussian source.

Via the analysis of $\MSE_{\sf LB}(\sigma^2)$, we obtain immediately the following necessary condition on the number of random linear measurements for the true \ac{MMSE} to approach zero. 
\begin{theorem}
\label{theo:2}
Consider the linear measurement model in (\ref{eq:model}) where $\mathbf{x} \sim \sum_k p_k \, \mathcal{N}( \boldsymbol{\mu}_{\mathbf{x}}^{(k)},\mathbf{\Sigma}_{\mathbf{x}}^{(k)})$, $s\sub{max}= \max_k s_k$, with $s_k= \rank (\mathbf{\Sigma}_{\mathbf{x}}^{(k)})$, and $\ell$ is the number of measurements. Then, with probability 1, it follows that: 
\begin{equation}
\lim_{\sigma^2 \rightarrow 0} \MMSE^{\sf GM}(\sigma^2) = 0   \Rightarrow  \ell \geq s\sub{max}.  
\end{equation}
\end{theorem}

\begin{IEEEproof}
It is evident that if $\MMSE^{\sf G}_k(\sigma^2) \to 0$ as $\sigma^2 \to 0$, for all $k$, then $\MSE_{\sf LB}(\sigma^2) \to 0$ as $\sigma^2 \to 0$. Theorem~\ref{theo:Gauss} proves that a necessary and sufficient condition for $\lim_{\sigma^2 \to 0}  \MMSE^{\sf G}_k(\sigma^2) = 0$ to hold with probability 1 is that $\ell \geq s_k$; therefore, a necessary condition for $\lim_{\sigma^2 \to 0} \MMSE^{\sf GM} (\sigma^2) = 0$ to hold with probability 1 is that $\ell \geq s\sub{max} = \max_ks_k$.
\end{IEEEproof}

An immediate corollary is the following first-order, low-noise expansion for the lower-bound of the \ac{MMSE} for \ac{GMM} inputs: 
\begin{equation}
\MSE_{\sf LB}(\sigma^2) =  \sum_k p_k \, \mathcal{M}^{\sf G}_{\infty_k} + \left( {\sum_k p_k \, \mathcal{D}_k^{\sf G }}\right) \sigma^2 + { o}\left( \sigma^2 \right),
\label{eq:lowexp}
\end{equation}
where $\mathcal{M}^{\sf G}_{\infty_k}$ and $\mathcal{D}_k^{\sf G }$ are the zero and first-order terms of the expansion of the Gaussian MMSE corresponding to class~$k$.


\subsection{Sufficient condition}
\label{par:suff}

The sufficient condition on the number of random linear measurements for the \ac{MMSE} phase transition to occur is based instead on the analysis of an upper bound to the \ac{MMSE}. We construct such an \ac{MMSE} upper bound -- denoted as $\MSE_{\sf CR}(\sigma^2)$ -- by using a (sub-optimal) \emph{classify and reconstruct} procedure\footnote{Note that the classify and reconstruct procedure is presented here as a mathematical tool to determine an upper bound to the number of measurements needed to obtain perfect reconstruction when using the optimal conditional mean estimator.}:
\begin{enumerate}
\item First, we obtain an estimate of the signal class by the \ac{MAP} classifier as follows:
\begin{equation}
\hat{c} = \arg \max_k p(\mathbf{y} | c=k ) p_k,
\end{equation}
where the variable $\hat{c}$ represents the estimate of the signal class, the random variable $c$ represents the actual signal class and $p(\mathbf{y}|c=k)$ denotes the conditioned \ac{pdf} of the measurement vector $\mathbf{y}$ given the signal class $k$;

\item Then, we recontruct the source vector $\mathbf{x}$ from the measurement vector $\mathbf{y}$ by using the conditional mean estimator associated with the class estimate $\hat{c}$ as follows:
\begin{equation}
\hat{\mathbf{x}}(\mathbf{y},c=\hat{c}) = \mathcal{W}_{\hat{c}}(\mathbf{y})  = \boldsymbol{\mu}_{\mathbf{x}}^{(\hat{c})}  +    \mathbf{W}_{\hat{c}} (\mathbf{y}  - \mathbf{\Phi} \boldsymbol{\mu}_{\mathbf{x}}^{(\hat{c})}).
\label{eq:WienerEst}
\end{equation}
\end{enumerate}
Note that a similar approach has been shown to offer state-of-the-art performance in the reconstruction of signals drawn from \ac{GMM} sources from compressive measurements~\cite{YuSapiro12,DuarteC13}.


The optimality of the conditional mean estimator together with the (in general) sub-optimality of the classify and reconstruction approach leads immediately to the fact that:
\begin{equation}
\MMSE^{\sf GM}(\sigma^2) \leq \MSE_{\sf CR}(\sigma^2).
\end{equation}

The analysis of the \emph{classify and reconstruct} \ac{MMSE}, which is aided by recent results on the characterization of the performance of \ac{GMM} classification problems from noisy compressive measurements~\cite{Reboredo13}, then leads to the following sufficient condition on the number of random linear measurements for the true \ac{MMSE} to approach zero.


\begin{theorem}
\label{theo:3}
Consider the linear measurement model in (\ref{eq:model}) where $\mathbf{x} \sim \sum_k p_k \, \mathcal{N}( \boldsymbol{\mu}_{\mathbf{x}}^{(k)},\mathbf{\Sigma}_{\mathbf{x}}^{(k)})$, $s\sub{max}= \max_k s_k$, with $s_k= \rank (\mathbf{\Sigma}_{\mathbf{x}}^{(k)})$, and $\ell$ is the number of measurements. Then, with probability 1, it holds: 
\begin{equation}
\ell >  s\sub{max}   \Rightarrow  \lim_{\sigma^2 \rightarrow 0} \MMSE^{\sf GM}(\sigma^2) = 0.   
\end{equation}
\end{theorem}

\begin{IEEEproof}
The proof of this Theorem is provided in Appendix~\ref{app:UpGMM}.
\end{IEEEproof}


\subsection{The \ac{MMSE} phase transition}

The characterization of the \ac{MMSE} phase transition follows by combining the results encapsulated in Theorems~\ref{theo:2} and \ref{theo:3}. In particular, it is possible to construct a sharp characterization of the transition that is accurate within one measurement where:
\begin{itemize}
\item When $\ell < s\sub{max}$, the function $\MMSE^{\sf GM}(\sigma^2)$ converges to an error floor as $\sigma^2 \to 0$;
\item When $\ell > s\sub{max}$, the function $\MMSE^{\sf GM}(\sigma^2)$ converges to zero as $\sigma^2 \to 0$; 
\item When $\ell = s\sub{max}$, the function $\MMSE^{\sf GM}(\sigma^2)$ may or may not approach zero as $\sigma^2 \to 0$, depending on the exact class dependent source covariances (see Section \ref{par:numres}). 
\end{itemize}

Note that -- akin to the Gaussian result -- one requires the number of linear random measurements to be greater than the largest of the dimensions of the subspaces spanned by the class dependent source covariances for the phase transition to occur. This is due to the fact that -- as reported in Appendix~\ref{app:UpGMM} -- with such a number of measurements one is able to classify perfectly and thereby to reconstruct perfectly in the low-noise regime. Note also that the \emph{classify and reconstruct} procedure is nearly ``phase transition'' optimal: the number of measurements required by such a procedure differs at most by one measurement from the number of measurements required by the optimal conditional mean estimation strategy. This also provides a rationale for the state-of-the-art results reported in~\cite{YuSapiro12,DuarteC13}, which are based on the use of the class conditioned Wiener filters for reconstruction and the detection of the \emph{a posteriori} most probable class of the signal.


\section{From random to designed kernels}
\label{sec:Design}

The emphasis of Sections \ref{par:Gaussinputs} and \ref{par:GMMinputs} has been on the derivation of necessary and sufficient conditions on the number of linear random measurements for the \ac{MMSE} phase transition to occur. However, in view of recent interest on the design of linear measurements in the literature~\cite{Ashok08,Baheti09,DuarteC12ICASSP,Carson12,Chen12}, it is also natural to ask whether designed kernels have an impact on such bounds.

In particular, we seek to characterize the impact on the \ac{MMSE} phase transition of kernels designed via the following optimization problem:
\begin{equation}
\label{optimization_problem}
\begin{aligned}
& \underset{\mathbf{\Phi}}{\text{minimize}}
& &  \MMSE(\sigma^2, \mathbf{\Phi})  \\
& \text{subject to}
& &  \tr \left( \mathbf{\Phi} \mathbf{\Phi}^{\dag} \right) \leq \ell
\end{aligned}
\end{equation}
where the constraint guarantees that, on average,  the rows of the designed kernel have unit $\ell_2$-norm. 
Note that, for the sake of clarity, we now express explicitly that the \ac{MMSE} is a function of the linear measurement kernel.

We start by showcasing the optimal linear kernel design for a Gaussian source, where ``optimality'' is defined by the optimization problem in (\ref{optimization_problem}), which follows immediately by leveraging results on the joint optimization of transmitter and receiver for coherent \ac{MIMO} transmission~\cite{Palomar03}.

\begin{theorem}
\label{optimal_kernel_Gauss}
Consider the linear measurement model in (\ref{eq:model}) where $\mathbf{x} \sim \mathcal{N}(\boldsymbol{\mu}_{\mathbf{x}}, \mathbf{\Sigma}_{\mathbf{x}})$, $s= \rank (\mathbf{\Sigma}_\mathbf{x})$ and $\ell$ is the number of measurements. Then, the measurement kernel $\mathbf{\Phi}^{\star}$ that solves the optimization problem in \eqref{optimization_problem} can be expressed as follows:
\begin{equation}
\mathbf{\Phi}^{\star} = \left[ \mathrm{diag} \left(  \sqrt{\lambda_{\mathbf{\Phi},1}^{\star}},  \ldots, \sqrt{\lambda_{\mathbf{\Phi},\ell}^{\star}}  \right)  \mathbf{0}_{\ell \times (n-\ell)}   \right]  \mathbf{U}_{\mathbf{x}}^\dag,
\label{eq:PhioptGauss}
\end{equation}
where the squared singular values of $\mathbf{\Phi}^{\star}$ are obtained through the water-filling principle~\cite{Cover91} as follows
\begin{equation}
\lambda_{\mathbf{\Phi},i}^{\star}  = \left[   \eta  -  \frac{\sigma^2}{\lambda_{\mathbf{x},i}} \right]^+,
\label{eq:PwrAll}
\end{equation}
and $\eta >0 $ is such that $\sum_{i=1}^\ell  \lambda_{\mathbf{\Phi},i}^{\star} \leq \ell$ and $[x]^+=\max \{  x,0  \}$. 
\end{theorem}

\begin{IEEEproof}
The problem of finding the projection matrix which minimizes the reconstruction {MMSE} for Gaussian input signals can be mapped, with appropriate modifications, to the problem of finding the linear precoder which minimizes the {MMSE} of a \ac{MIMO} transmission system. In particular, the object function of this minimization problem is a Schur-concave function of the \ac{MMSE} matrix and the optimal linear precoder is shown to diagonalize the \ac{MIMO} channel matrix~\cite[Theroem 1]{Palomar03}. This implies that, in our scenario, the designed kernel right singular vectors correspond to the source covariance eigenvectors, i.e., that the measurement kernel that minimizes the MMSE exposes the modes of the source covariance\footnote{In general, when the additive Gaussian noise $\mathbf{w}$ has a non diagonal covariance matrix $ \mathbf{\Sigma}_{\mathbf{w}}$, the measurement kernel $\mathbf{\Phi}^{\star}$ can be shown to expose and align the modes of both the input source and the noise, as observed for the measurement kernel which maximizes the mutual information between $\mathbf{x}$ and $\mathbf{y}$~\cite{Carson12}.}. 
Then, the fact that the squared singular values of $\mathbf{\Phi}^\star$ obey the water-filing type of interpretation in (\ref{eq:PwrAll}) follows from the \ac{KKT} conditions associated with the optimization problem yielded by taking the kernel right singular vectors to correspond to the source covariance eigenvectors.
\end{IEEEproof}

It is now straightforward to show that kernel design does not impact the phase transition of the \ac{MMSE} associated with Gaussian sources. However, and despite the fact that kernel design does not affect the number of measurements necessary to observe the phase transition, we also show that there is value in using designed kernels \emph{in lieu} of random ones because one can thus improve reconstruction performance in terms both of a lower error floor (if present) and a lower power offset.

\begin{theorem}
\label{theo:GaussDes}
Consider the linear measurement model in (\ref{eq:model}) where $\mathbf{x} \sim \mathcal{N}(\boldsymbol{\mu}_{\mathbf{x}}, \mathbf{\Sigma}_{\mathbf{x}})$, $s= \rank (\mathbf{\Sigma}_\mathbf{x})$, $\ell$ is the number of measurements and $\mathbf{\Phi} =\mathbf{\Phi}^{\star}$, where $\mathbf{\Phi}^{\star}$ solves the optimization problem in \eqref{optimization_problem}. 
Then, the first-order, low-noise expansion of the \ac{MMSE} is given by:
\begin{equation}
\MMSE^{\sf G}(\sigma^2,\mathbf{\Phi}^{\star})  = \mathcal{M}_{\infty}^{\sf GD} +  \mathcal{D}^{\sf GD} \cdot \sigma^2 + o(\sigma^2).
\label{eq:expGAD}
\end{equation}
where
\begin{equation}
\mathcal{M}_{\infty}^{\sf GD} =  \sum_{i=\ell'+1}^s \lambda_{\mathbf{x},i}  \qv \mathcal{D}^{\sf GD}= (\ell')^2/\ell
\end{equation}
and $\ell' = \min\{s,\ell \}$.
\end{theorem}
\begin{IEEEproof}
On substituting the expression of $\mathbf{\Phi}^{\star}$ in (\ref{eq:PhioptGauss}) into (\ref{eq:MMSEGauss}), it is possible to expand the \ac{MMSE} associated with the optimal linear kernel design as follows: 
\begin{equation}
\MMSE^{\sf G}(\sigma^2,\mathbf{\Phi}^{\star})   = \sum_{i=1}^{\ell'} \frac{\lambda_{\mathbf{x},i}}{1  +  \frac{1}{\sigma^2} \lambda_{\mathbf{x},i} \lambda_{\mathbf{\Phi},i}^{\star} } + \sum_{i=\ell'+1}^s \lambda_{\mathbf{x},i}.
\label{eq:MMSEGDes}
\end{equation}
Observe that in the limit $\sigma^2 \to 0$, it follows from (\ref{eq:PwrAll}) that $\lambda_{\mathbf{\Phi},i}^{\star}= \ell / \ell'$ for $i=1,\ldots,\ell'$ and $\lambda_{\mathbf{\Phi},i}^{\star}= 0$ for $i=\ell',\ldots,\ell$.
Moreover, notice also that the second term in (\ref{eq:MMSEGDes}) is identically equal to zero if and only if $\ell \geq s$. 
\end{IEEEproof}

Note that when $\ell < s$ the error floor corresponds to the source power that the kernel fails to capture in view of its compressive nature. On the other hand, when $\ell \geq s$ the \ac{MMSE} decay associated with an optimal kernel is equal to that of a random one, i.e. $\mathcal{O} (\sigma^2)$, but the \ac{MMSE} power offset is lower: it is only a function of the number of measurements and the dimension of the subspace spanned by the source, independently of the exact form of the eigenvectors or eigenvalues of the source covariance.

Finally, it is also straightforward to show that kernel design does not impact the phase transition of the \ac{MMSE} associated with \ac{GMM} sources. The minimum number of measurements required to perfectly reconstruct the signal in the low-noise regime is also $\ell > s\sub{max}$ in this case. However, careful kernel design can increase the system performance by guaranteeing lower error floors and power offsets. 

The method leveraged to prove this result is also based on the analysis of lower and upper bounds to the \ac{MMSE} associated with the reconstruction of signals drawn from a \ac{GMM} source that are sensed now via the optimal linear kernel design. 
In particular, we consider a lower bound -- which we denote by $\MSE_{\sf LBD}(\sigma^2)$ -- which is expressed in terms of the value of the \ac{MMSE} corresponding to the actual Gaussian class from which each realization of the input signal is drawn and the corresponding optimal kernel $\mathbf{\Phi}^{\star}_k$ in (\ref{eq:PhioptGauss}). Then,
\begin{IEEEeqnarray}{rCl}
\MMSE^{\sf GM}(\sigma^2, \mathbf{\Phi}^\star) & \geq & \MSE_{\sf LBD}(\sigma^2) \\
&=& \sum_k p_k  \,  \MMSE_k^{\sf G} (\sigma^2,\mathbf{\Phi}^{\star}_k).
\label{eq:LBgenie}
\end{IEEEeqnarray}
We also consider a trivial upper bound: the MMSE associated with the optimal kernel design can always be upper bounded by the MMSE associated with a random kernel design.



\begin{theorem}
\label{theo:4}
Consider the linear measurement model in (\ref{eq:model}) where $\mathbf{x} \sim \sum_k p_k \, \mathcal{N}( \boldsymbol{\mu}_{\mathbf{x}}^{(k)},\mathbf{\Sigma}_{\mathbf{x}}^{(k)})$, $s\sub{max}= \max_k s_k$, with $s_k= \rank (\mathbf{\Sigma}_{\mathbf{x}}^{(k)})$, $\ell$ is the number of measurements and $\mathbf{\Phi}=\mathbf{\Phi}^{\star}$, where $\mathbf{\Phi}^{\star}$ that solves the optimization problem in \eqref{optimization_problem}. Then, it holds 
\begin{equation}
\label{eq:nec_des}
\lim_{\sigma^2 \rightarrow 0} \MMSE^{\sf GM}(\sigma^2,\mathbf{\Phi}^\star) = 0   \Rightarrow  \ell \geq s\sub{max},  
\end{equation}
and 
\begin{equation}
\label{eq:suff_des}
\ell >  s\sub{max}   \Rightarrow  \lim_{\sigma^2 \rightarrow 0} \MMSE^{\sf GM}(\sigma^2,\mathbf{\Phi}^\star) = 0.   
\end{equation}
\end{theorem}

\begin{IEEEproof}
It is possible to prove the sufficient condition (\ref{eq:suff_des}) by observing that 
\begin{equation}
\MMSE^{\sf GM}(\sigma^2,\mathbf{\Phi}^\star) \leq \MMSE^{\sf GM}(\sigma^2,\mathbf{\Phi})
\end{equation}
for all possible measurement matrices $\mathbf{\Phi}$ that verify the trace constraint in (\ref{optimization_problem}). Among them, we can consider random kernels with Gaussian, zero-mean and fixed-variance, \ac{i.i.d.} entries and we can obtain the sufficient condition by leveraging directly the result in Theorem~\ref{theo:3}. On the other hand, in order to prove the necessary condition (\ref{eq:nec_des}),  consider the lower bound in (\ref{eq:LBgenie}) and observe that
\begin{equation}
\MMSE^{\sf G}_k(\sigma^2,\mathbf{\Phi}^\star_k) \leq \MMSE^{\sf G}_k(\sigma^2,\mathbf{\Phi})
\end{equation}
for all possible measurement matrices $\mathbf{\Phi}$ that verify the trace constraint in (\ref{optimization_problem}). Also in this case, we can consider random kernels with Gaussian, zero-mean and fixed-variance, \ac{i.i.d.} entries. Then, it is possible prove (\ref{eq:nec_des}) by leveraging the necessary condition embedded in Theorem~\ref{theo:GaussDes}, that is,
\begin{equation}
\lim_{\sigma^2 \to 0} \MMSE^{\sf G}_k(\sigma^2,\mathbf{\Phi}) =0 \Rightarrow \ell \geq s_k.
\end{equation}
\end{IEEEproof}



In view of (\ref{eq:expGAD}) and (\ref{eq:LBgenie}), the  low-noise expansion of the proposed lower bound is given by:
\begin{equation}
 \MSE_{\sf LBD}(\sigma^2) = \sum_k p_k  \mathcal{M}_{\infty_k}^{\sf GD} + \left( \sum_k p_k \mathcal{D}^{\sf GD}_k\right)  \sigma^2 + o(\sigma^2),
\end{equation}
where $\mathcal{M}^{\sf GD}_{\infty_k}$ and $\mathcal{D}_k^{\sf GD }$ are the zero-order term and the coefficient of the first-order term of the expansion of the Gaussian MMSE corresponding to class~$k$. We also conclude that, though kernel design may not increase the \ac{MMSE} decay rate beyond $\mathcal{O}(\sigma^2)$, it can have an impact on the \ac{MMSE} power offset associated with \ac{GMM} sources. 

It is important to note though that this analysis has concentrated on offline designs where the $\ell$ measurements are designed concurrently~\cite{Carson12,DuarteC12ICASSP}, rather than online kernel designs that entail a sequential design of the measurements by leveraging information derived from previous measurements~\cite{DuarteC11,Carson12,DuarteC13}. It is possible that such online designs have an impact on the phase transition.

\section{Numerical results}
\label{par:numres}

We now provide results both with synthetic and real data to illustrate the theory. Recovery is based upon the conditional expectation, which is analytic for the case of GMM priors, and optimal in terms of mean-squared error. In all simulations we use random measurement kernels, where the entries of the kernel are realizations of \ac{i.i.d.} Gaussian random variables with zero mean and unit variance which are subsequently normalized by a scaling factor that guarantees that $\tr \left( \mathbf{\Phi}^\dag \mathbf{\Phi} \right)  \leq \ell$.

\subsection{Synthetic data}


\begin{figure}
\begin{center}
\includegraphics[width=0.53\textwidth]{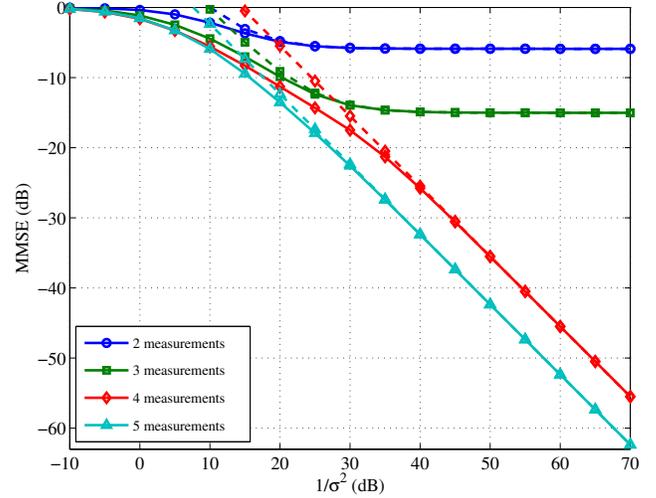}
\caption{\ac{MMSE} vs. $1/\sigma^2$ for different numbers of random measurements $\ell=2,3,4,5$ for a Gaussian source with $n=5$ and $s=4$. Actual \ac{MMSE} (solid lines) and low-noise, first-order expansions (dashed lines).}
\label{fig:Gauss}
\end{center}
\end{figure}

Fig.~\ref{fig:Gauss} shows the \ac{MMSE}  vs. $1/\sigma^2$ for a Gaussian source with dimension $n=5$ and rank $s=4$. We confirm that the \ac{MMSE} phase transition occurs with $\ell = 4$ measurements. We also confirm that the first-order expansion in (\ref{eq:expGauss}) captures well the behavior of the \ac{MMSE}, both in the presence and absence of an error floor, for values of $1/\sigma^2$ larger than 20-30 dB. In fact, such noise amplitudes are already well below $1/\sigma^2=60$\, dB, which is a noise level with operational significance for various image processing applications~\cite{Carson12,Chen12}. Note also that by taking the number of measurements to be greater than the number of measurements that achieves the phase transition we do not affect the \ac{MMSE} decay but we only affect the \ac{MMSE} power offset.


\begin{figure}
\begin{center}
\includegraphics[width=0.53\textwidth]{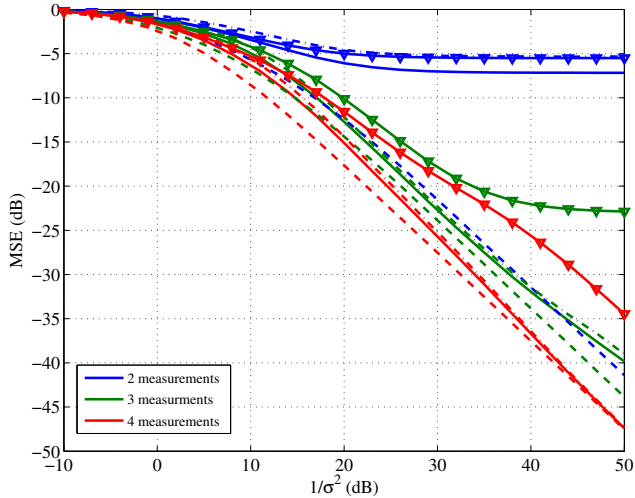}
\caption{\ac{MMSE} vs. $1/\sigma^2$ for different numbers of random measurements $\ell = 2,3,4$ for a 2-classes GMM source with $s_1=s_2=2$. Actual MMSE (solid lines), lower bound (dashed lines), CR upper bound (dashed-dotted lines) and LMMSE upper bound (triangles).}
\label{fig_GMM1}
\end{center}
\end{figure}


Fig.~\ref{fig_GMM1} now shows the values of the \ac{MMSE} for a 2-classes \ac{GMM} input with $n=4$, $p_1=p_2=0.5$, means $\boldsymbol{\mu}_{\mathbf{x}}^{(1)}=\boldsymbol{\mu}_{\mathbf{x}}^{(2)}=\mathbf{0}$, and covariances drawn from a central Wishart distribution~\cite[p. 84]{Srivastava79} with dimension 4 and degrees of freedom 2, so that $s_1=s_2=2$. 
We report the actual value of $\MMSE^{\sf GM}(\sigma^2)$, the lower bound $\MSE_{\sf LB}(\sigma^2)$, and the classify and reconstruct upper bound $\MSE_{\sf CR}(\sigma^2)$. We also report another \ac{MMSE} upper bound associated with a linear estimator, i.e.,  the \ac{LMMSE}~\cite{Flam11}. We notice that when $\ell =2$ the lower bound converges to zero as $\sigma^2 \to 0$, whereas the classify and reconstruct upper bound converges to an error floor. However, it appears that the classify and reconstruct upper bound captures better the features of the actual \ac{MMSE}, which also exhibits an error floor. On the other hand, notice that when $\ell \geq 3$  both the lower and upper bound converge to zero as $\sigma^2 \to 0$, as expected. It is also interesting to observe that the \ac{LMMSE} upper bound does not describe crisply the \ac{MMSE} phase transition: it is in fact possible to show by leveraging the previous machinery that such a sub-optimal estimator requires the number of random measurements to be larger than or equal to the dimension of the direct sum of the subspaces spanned by the different signals in the different classes for a \ac{MMSE} phase transition to occur.

Another interesting feature relates to the fact that when $\ell = 3$ both the upper bound to the \ac{MMSE} and the actual \ac{MMSE} are $\mathcal{O} (\sigma)$ as $\sigma^2 \to 0$ whereas when $\ell > 3$ the upper and lower bound to the \ac{MMSE} and the actual \ac{MMSE} are $\mathcal{O} (\sigma^2)$ as $\sigma^2 \to 0$. This behavior appears to be related to the fact that when $\ell = 3$ or $\ell > 3$ the misclassification probability of the optimal \ac{MAP} classifier, which is also used in the classify and reconstruction procedure, approaches zero as $\sigma^2 \to 0$ with different decays~\cite{Reboredo13}. This behavior stands in contrast to the scalar case~\cite{Wu11} and implies that the \ac{MMSE} dimension corresponding to the mixture of Gaussian vectors is not equal to the mixture of the \ac{MMSE} dimensions of the individual Gaussian vectors: this is only true when the first-order expansions of lower and the upper bound coincide.

%



\subsection{Real data}
\label{par:RealData}

The phase transition phenomena can also be observed in the reconstruction of real imagery data. As an example, we consider a $256 \times 256$ cropped version of the image ``barbara''.  
The input $\mathbf{x}$, in this case, represents $8 \times 8$ non-overlapping patches extracted from the image.  The source is described by a 20-classes \ac{GMM} prior that is obtained by training the non-parametric, Bayesian, dictionary learning algorithm described in~\cite{Chen10} over 100,000 patches randomly extracted from 500 images in the Berkeley Segmentation Dataset\footnote{http://www.eecs.berkeley.edu/Research/Projects/CS/vision/grouping/\\resources.html}. Note that the image ``barbara'' is not in the training ensemble. The so obtained \ac{GMM} prior has full rank, class conditioned input covariance matrices $\mathbf{\Sigma}_{\mathbf{x}}^{(k)}$. In order to fit the trained \ac{GMM} to the low-rank model (or, equivalently, to the model of a union of subspaces), which is the basis of our theory, only the first $s\sub{max} = 14$ principal components of each class-conditioned input covariance matrix are retained, and the remaining $50$ eigenvalues of each covariance matrix are set to be equal to zero. 
Moreover, our test image is modified by projecting each patch extracted from the image ``barbara'' onto the $14$-dimensional sub-space corresponding to the low-rank input covariance matrix of the class associated with that particular patch. 
Note that projecting the image onto the lower dimensional subspaces does not introduce substantial distortion. In fact, the \ac{PSNR} of the projected image with respect to the original ground truth in this case is equal to $77.3$\,dB\footnote{The \ac{PSNR} values obtained by choosing $s\sub{max} = 13$ and $s\sub{max} = 15$ are $76.9$\,dB and $78$\,dB, respectively. On the other hand, setting $s\sub{max} =10$ reduces the \ac{PSNR} to $75.2$\,dB and it induces visible distortion effects.}. This is a manifestation of the fact that natural images are well represented by ``almost low-rank'' \ac{GMM} priors, and the eigenvalues of the corresponding class conditioned input covariance matrices decay rapidly. This underscores that the low-rank GMM representation is a good model for patches extracted from natural imagery, and therefore of significant practical value.

Reconstruction of the vectors $\mathbf{x}$ from the compressive measurements $\mathbf{y}$ is performed by the conditional mean estimator corresponding to the trained \ac{GMM} prior after taking the 14 principal components, which can be written in closed-form~\cite{Chen10}.

\begin{figure}
\begin{center}
\includegraphics[width=0.53\textwidth]{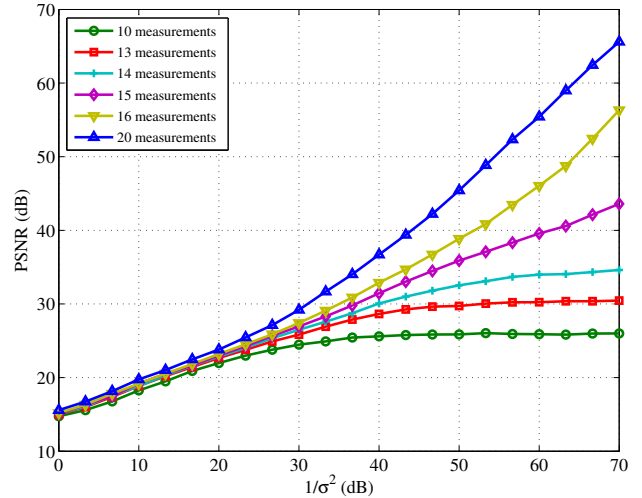}
\caption{PSNR vs. $1/\sigma^2$. Image ``barbara.png'', projected onto a \ac{GMM} model with $s\sub{max}=14$.}
\label{fig_BarbaraPSNR}
\end{center}
\end{figure}

\begin{figure*}
\begin{center}
\begin{tabular}{cccc}
\includegraphics[width=0.225\textwidth]{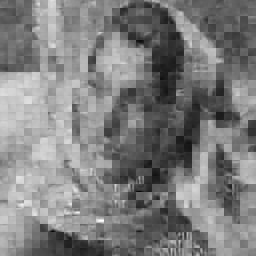} &
\includegraphics[width=0.225\textwidth]{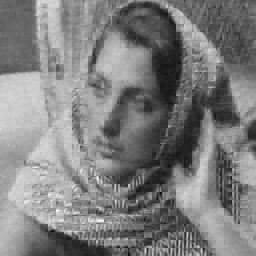} &
\includegraphics[width=0.225\textwidth]{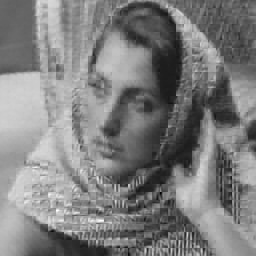} & 
\includegraphics[width=0.225\textwidth]{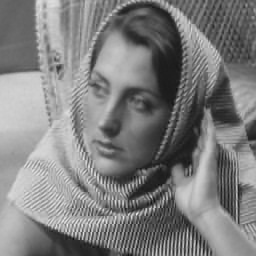}\\
\includegraphics[width=0.225\textwidth]{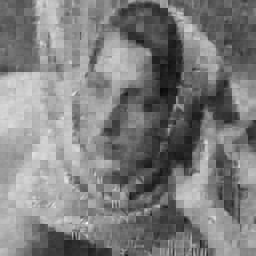} &
\includegraphics[width=0.225\textwidth]{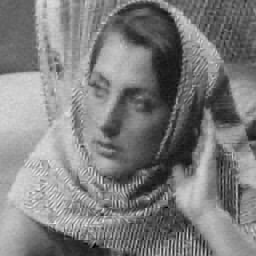} &
\includegraphics[width=0.225\textwidth]{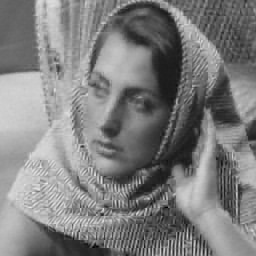} & 
\includegraphics[width=0.225\textwidth]{barbara14.png}\\
\includegraphics[width=0.225\textwidth]{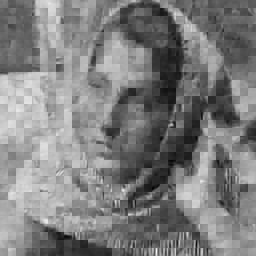} &
\includegraphics[width=0.225\textwidth]{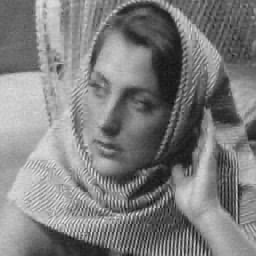} &
\includegraphics[width=0.225\textwidth]{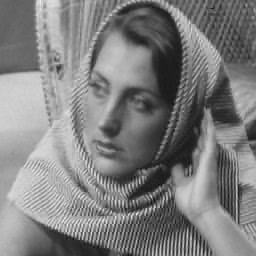} & 
\includegraphics[width=0.225\textwidth]{barbara14.png}\\
\end{tabular}
\end{center}
\caption{Barbara. The right hand column contains the non-compressed image ``barbara.png'' after projection onto a \ac{GMM} model with $s\sub{max}=14$. The three rows correspond to $\ell = 8, 12, 16$ random measurements, from the top to the bottom. The first three columns, from left to right, correspond to the noise levels $1/\sigma^2 = 20,40, 60$\, dB.}
\label{fig_BarbaraExamples}
\end{figure*}

Fig.~\ref{fig_BarbaraPSNR} shows the \ac{PSNR} vs. $1/\sigma^2$ for different numbers of compressive measurements.  It is possible to clearly observe the phase transition when $\ell > s\sub{max}$ and that, similarly to what has been noted in Fig.~\ref{fig_GMM1}, the \ac{PSNR}  increases approximately as $\mathcal{O}(1/\sigma)$ when $\ell =15$, and as $\mathcal{O}(1/\sigma^2)$ when $\ell > 15$. Finally, in Fig.~\ref{fig_BarbaraExamples} are reported some reconstruction examples. The right hand column contains the image ``barbara'' projected onto the union of 14-dimensional subspaces that characterize the \ac{GMM} prior. The three rows correspond to $\ell = 8, 12, 16$ random measurements, from the top to the bottom. The first three columns, from left to right, correspond to the noise levels $1/\sigma^2 = 20,40, 60$\, dB. When the number of measurements is below the phase transition, perfect reconstruction is not possible, even when $\sigma^2 \to 0$: this is particularly evident by observing the high-frequency content of the image. On the other hand, when $\ell = 16$, a clear phase transition is observable.

%

It is also relevant to reflect further on the fact that our theory applies only to low-rank rather than the so-called ``approximately'' low-rank models. Arguably, a \ac{GMM} model trained with natural images is not exactly low-rank, as real signals  do not perfectly lie on the union of low-dimensional subspaces~\cite{Chen10}. 
In fact, it is typical to describe the class conditioned covariance matrices associated with a \ac{GMM} model as follows:
\begin{equation}
\mathbf{\Sigma}_{\mathbf{x}}^{(k)} =  \bar{\mathbf{\Sigma}}_{\mathbf{x}}^{(k)}  + \epsilon \, \mathbf{I}_{n},
\end{equation}
where the matrix $\bar{\mathbf{\Sigma}}_{\mathbf{x}}^{(k)}$ is exactly low-rank, and the matrix $\epsilon \, \mathbf{I}_{n}$ accounts for model mismatch between real data and their projection onto the principal components that contain the large majority of the information associated to the data. Given the fact that the compressive sensing model in (\ref{eq:model}) with $\mathbf{w} \sim \mathcal{N}(\mathbf{0}, \sigma^2 \, \mathbf{I})$ and full-rank GMM with class conditioned covariances $\mathbf{\Sigma}_{\mathbf{x}}^{(k)}$ is mathematically equivalent to the compressive sensing model in (\ref{eq:model}) with $\mathbf{w} \sim \mathcal{N}(\mathbf{0}, \epsilon \, \mathbf{\Phi}\mathbf{\Phi}^{\dag}  +  \sigma^2 \, \mathbf{I})$ and low-rank GMM with class conditioned covariances $ \bar{\mathbf{\Sigma}}_{\mathbf{x}}^{(k)} $, then it is possible to appreciate the impact of model mismatch on the theory.

For example, consider our earlier testing model with images that are not projected onto lower-dimensional union of subspaces. It is evident that the reconstruction \ac{PSNR} would now be upper bounded as $\sigma^2 \to 0$ for all  $\ell < n$, in view of the noise amplification from $\sigma^2$ to roughly\footnote{In fact, the noise power depends also on the value of $\tr(\mathbf{\Phi}\mathbf{\Phi}^{\dag})$. However, when the entries of $\mathbf{\Phi}$ are \ac{i.i.d.}, zero-mean, Gaussian, the matrix $\mathbf{\Phi}\mathbf{\Phi}^{\dag}$ approximates well the identity matrix.} $1/(\sigma^2 + \epsilon)$. This also leads to the conclusions that the performance of the ``approximately'' low-rank model as $\sigma^2 \to 0$ is comparable to the performance of the low-rank model with $1/\sigma^2 = 1/\epsilon$. That is, operating an ``approximately'' low-rank model at a certain $\sigma^2$ leads to a performance that is comparable to that of operating a low-rank model at $\sigma^2 + \epsilon$.

Nonetheless, it is worth noting that natural images projected on low-rank union of subspaces turn out to be very good approximations of the original images, as it was shown for the case of the image ``barbara'' (see Fig.~\ref{fig_BarbaraExamples}), thus validating the effectiveness of this model in representing real data.

\section{Conclusion}
\label{par:conclusion}

The principal contribution is a non-asymptotic (in the number of dimensions) characterization of \ac{MMSE} phase transitions associated with the reconstruction of \ac{GMM} signals from noisy compressive measurements. In particular, it has been shown that with either random or optimal kernel designs it is sufficient to take the number of measurements to be strictly greater than the largest dimension of the sub-spaces spanned by the class-conditioned signals in the \ac{GMM} model -- a dual of sparsity -- for a phase transition to occur. It has also been shown that additional measurements and/or designed measurements translate only onto (occasionally) an improved \ac{MMSE} decay or an improved \ac{MMSE} power offset in the low-noise regime. Another interesting by-product of the contribution is the fact that a (sub-optimal) classify and reconstruction procedure is nearly phase-transition optimal. This also provides a rationale for the state-of-the-art performance of similar reconstruction procedures, as reported in the recent literature~\cite{YuSapiro12,DuarteC13}.

\appendices

\section{Useful lemmas}
\label{app:lemmas}
\begin{lemma}
\label{lem:1}
Let $\mathbf{A} \in \mathbb{R}^{n \times n}$ be a positive semidefinite matrix. Then, for all $\mathbf{x} \in \mathbb{R}^n$,
\begin{equation}
\mathbf{A} \mathbf{x} = \mathbf{0}   \Leftrightarrow   \mathbf{x}^\dag  \mathbf{A} \mathbf{x} = 0.
\label{eq:lemnull}
\end{equation}
\end{lemma}

\begin{IEEEproof}
If $\mathbf{x}^\dag  \mathbf{A} \mathbf{x} = 0$, then, $\|  \mathbf{A}^{\frac{1}{2}} \mathbf{x} \|^2 =0$, in which $ \mathbf{A}^{\frac{1}{2}}$ is the positive semidefinite matrix square root of $\mathbf{A}$. Therefore, $\mathbf{A}^{\frac{1}{2}} \mathbf{x}= \mathbf{0} $, which also implies $\mathbf{A} \mathbf{x}= \mathbf{0} $. The opposite implication in (\ref{eq:lemnull}) is straightforward.
\end{IEEEproof}

\begin{lemma}
\label{lem:overlap}
Given two positive semidefinite matrices, $\mathbf{A}_1, \mathbf{A}_2 \in \mathbb{R}^{n \times n}$, with ranks $s_1$ and $s_2$, respectively, and $s_{12} = \mathrm{rank}(\mathbf{A}_1 + \mathbf{A}_2)$, then
\begin{IEEEeqnarray}{rCl}
\nonumber
\frac{s_1 + s_2}{2} = s_{12} & \Leftrightarrow &\mathrm{Null}(\mathbf{A}_1) = \mathrm{Null}(\mathbf{A}_2) \\
& \Leftrightarrow & \mathrm{Im}(\mathbf{A}_1) = \mathrm{Im} (\mathbf{A}_2)
\label{eq:lemma}
\end{IEEEeqnarray}
\end{lemma}

\begin{IEEEproof}
First, it is easy to show that 
\begin{equation}
\mathrm{Null}(\mathbf{A}_1 +\mathbf{A}_2) = \mathrm{Null}(\mathbf{A}_1) \cap \mathrm{Null}(\mathbf{A}_2),
\label{eq:nullintersec}
\end{equation}
by leveraging Lemma~\ref{lem:1} and considering that $\mathbf{x}^\dag(\mathbf{A}_1 +\mathbf{A}_2)\mathbf{x} = 0$ if and only if $\mathbf{x}^\dag \mathbf{A}_1 \mathbf{x} = 0$ and  $\mathbf{x}^\dag \mathbf{A}_1 \mathbf{x} = 0$. Therefore, $\dim \mathrm{Null}(\mathbf{A}_1 +\mathbf{A}_2) \leq \dim  \mathrm{Null}(\mathbf{A}_1)$ and $\dim \mathrm{Null}(\mathbf{A}_1 +\mathbf{A}_2) \leq \dim  \mathrm{Null}(\mathbf{A}_2)$, which immediately implies $s_{12} \geq \max\{ s_1,s_2 \}$, and, in our case, $s_1 = s_2 =s_{12}$.
As a consequence, we have that
\begin{equation}
\dim \mathrm{Null}(\mathbf{A}_1) = \dim \mathrm{Null}(\mathbf{A}_2) = \dim \mathrm{Null}(\mathbf{A}_1 +\mathbf{A}_2),
\end{equation}
which can be combined with (\ref{eq:nullintersec}) to give
\begin{equation}
\mathrm{Null}(\mathbf{A}_1) = \mathrm{Null}(\mathbf{A}_2).
\end{equation}
Moreover, the last part of (\ref{eq:lemma}) can be easily obtained by observing that the image of a positive semidefinite matrix is the orthogonal complement of its null space.
\end{IEEEproof}

%

\begin{lemma}
\label{lem:rankim}
Let $\mathbf{A} \in \mathbb{R}^{n \times n}$ be a positive semidefinite matrix with $s=\rank (\mathbf{A})$ and let $\mathbf{x} \in \mathbb{R}^n$. Then,
\begin{equation}
\rank (\mathbf{A} + \mathbf{x}\mathbf{x}^\dag) = s + 1  \Leftrightarrow \mathbf{x} \notin \mathrm{Im}(\mathbf{A}).
\end{equation}
\end{lemma}

\begin{IEEEproof}
First, note that $\rank (\mathbf{A} + \mathbf{x}\mathbf{x}^\dag)  \leq s + 1$~\cite[\S 0.4.5.d]{Horn}. Then, the proof is based on expressing $\mathbf{A}$ in terms of its eigenvalues $\lambda_{\mathbf{A}_i}$ and eigenvectors $\mathbf{u}_{\mathbf{A}_i}$ as $\mathbf{A} =\sum_i \lambda_{\mathbf{A}_i}\mathbf{u}_{\mathbf{A}_i}\mathbf{u}_{\mathbf{A}_i}^\dag$.
\end{IEEEproof}

\section{Proof of Theorem~\ref{theo:Gauss}}
\label{app:Gauss}


We can rewrite the expression of the \ac{MMSE} for Gaussian input sources in (\ref{eq:MMSEGauss}) 
by making use of the cyclic property of the trace and the matrix inversion Lemma~\cite[\S 0.7.4]{Horn}
\begin{equation}
\mathbf{I}-\mathbf{A}^{\dag}(\mathbf{A}\mathbf{A}^{\dag}+c^{-1}\mathbf{I})^{-1}\mathbf{A}=(c\mathbf{I}+\mathbf{A}^{\dag}\mathbf{A})^{-1},
\end{equation}
with $\mathbf{A}=\mathbf{\Phi} \mathbf{\Sigma}_{\mathbf{x}}^{\frac{1}{2}}$, as
\begin{IEEEeqnarray}{rCl}
\MMSE^{\sf G}(\sigma^2)   & = &  \tr \left(  \mathbf{\Sigma}_{\mathbf{x}}  \left(  \mathbf{I} + 1/\sigma^2   \mathbf{\Sigma}_{\mathbf{x}}^{\frac{1}{2}} \mathbf{\Phi}^\dag \mathbf{\Phi} \mathbf{\Sigma}_{\mathbf{x}}^\frac{1}{2}   \right)^{-1}    \right) \\
& =  & \tr \left(  \mathbf{\Sigma}_{\mathbf{x}} \mathbf{U}  \left(  \mathbf{I} +  1/\sigma^2   \mathbf{\Lambda}   \right)^{-1}  \mathbf{U}^\dag   \right)  \\
& = & \tr \left(  \mathbf{\Sigma}_{\mathbf{x}} \mathbf{U}  \tilde{\mathbf{\Lambda}} \mathbf{U}^\dag   \right) ,
\end{IEEEeqnarray}
where we have used the eigenvalue decomposition of $\mathbf{\Sigma}$ in (\ref{eq:sigma}) and introduced the diagonal matrix $\tilde{\mathbf{\Lambda}} = \mathrm{diag}\left( \frac{1}{1+  \lambda_1/\sigma^2},\ldots,\frac{1}{1 +  \lambda_{\ell'}/\sigma^2},1,\dots,1 \right) $. Note now that the eigenvalue decomposition of a positive  semidefinite matrix can also be written as
\begin{equation}
\mathbf{A} = \mathbf{U}_{\mathbf{A}} \mathbf{\Lambda}_{\mathbf{A}} \mathbf{U}_{\mathbf{A}}^\dag = \sum_i \lambda_{\mathbf{A},i} \mathbf{u}_{\mathbf{A,i}}  \mathbf{u}_{\mathbf{A,i}}^\dag,
\end{equation}
in which $\mathbf{u}_{\mathbf{A,i}}$ is the $i$-th column of $\mathbf{U}_{\mathbf{A}}$. Therefore, the \ac{MMSE} for Gaussian inputs can be expressed as
\begin{equation}
\MMSE^{\sf G}(\sigma^2) =  \sum_{i=1}^{\ell'}  \frac{1}{1+  \lambda_i/\sigma^2} \mathbf{u}_i^\dag \mathbf{\Sigma}_{\mathbf{x}} \mathbf{u}_i   +  \sum_{i=\ell' + 1 }^n \mathbf{u}_i^\dag \mathbf{\Sigma}_{\mathbf{x}} \mathbf{u}_i 
\label{eq:MMSEsum}
\end{equation}
where $\mathbf{u}_1,\ldots,\mathbf{u}_{\ell'}$ are the eigenvectors of $\mathbf{\Sigma}$ corresponding to the positive eigenvalues $\lambda_1,\ldots, \lambda_{\ell'}$ and the vectors $\mathbf{u}_{\ell'+1},\ldots,\mathbf{u}_n$ can form any orthonormal basis of the null space $\mathrm{Null}(\mathbf{\Sigma})$, which has dimension $\dim ( \mathrm{Null}(\mathbf{\Sigma}))=n-\ell'$. The first term in (\ref{eq:MMSEsum}) tends to zero when $\sigma^2 \to 0$, so that the phase transition of the \ac{MMSE} is determined by the second term, which can be characterized on the basis of the description of the two null spaces $\mathrm{Null}(\mathbf{\Sigma})$ and $\mathrm{Null}(\mathbf{\Sigma}_{\mathbf{x}})$. In particular, note that 
\begin{equation}
\label{eq:contained}
\mathrm{Null}(\mathbf{\Sigma}_{\mathbf{x}})  \subseteq \mathrm{Null}(\mathbf{\Sigma}),
\end{equation}
as for each $\mathbf{v} \in \mathbb{R}^n$ such that  $\mathbf{\Sigma}_{\mathbf{x}} \mathbf{v}=\mathbf{0}$, it holds also $\mathbf{\Sigma} \mathbf{v}=\mathbf{0}$, since $\mathbf{\Sigma}^{\frac{1}{2}}_{\mathbf{x}} \mathbf{v}=\mathbf{0}$. 
Therefore, we can choose the basis  $\mathbf{u}_{\ell'+1},\ldots,\mathbf{u}_n$ as follows. The last $n-s$ vectors $\mathbf{u}_{s+1},\ldots,\mathbf{u}_n$ form an othonormal basis of of the space $\mathrm{Null}(\mathbf{\Sigma}_{\mathbf{x}})$ and the remaining vectors $\mathbf{u}_{\ell'+1},\ldots,\mathbf{u}_{s}$ form an orthonormal basis of $\mathrm{Null}(\mathbf{\Sigma})  \cap \mathrm{Null}(\mathbf{\Sigma}_{\mathbf{x}})^\perp$. Then, we can write (\ref{eq:MMSEsum}) as
\begin{equation}
\MMSE^{\sf G}(\sigma^2)  =   \sum_{i=1}^{\ell'}  \frac{1}{1+  \lambda_i / \sigma^2} \mathbf{u}_i^\dag \mathbf{\Sigma}_{\mathbf{x}} \mathbf{u}_i   +  \sum_{i=\ell' + 1 }^{s} {\mathbf{u}}_i^\dag \mathbf{\Sigma}_{\mathbf{x}} {\mathbf{u}}_i. 
\label{eq:MMSEsum2}
\end{equation}
Observe that, when $\ell \geq s$, then $\ell'=\ell\geq s$, and the second term in (\ref{eq:MMSEsum2}) is identically equally to zero, as the linear space $\mathrm{Null}(\mathbf{\Sigma})  \cap \mathrm{Null}(\mathbf{\Sigma}_{\mathbf{x}})^\perp$ contains only the zero vector, and the Gaussian \ac{MMSE} tends to zero when $\sigma^2 \to 0$. On the other hand, if $\ell < s$, by Lemma~\ref{lem:1}, $ {\mathbf{u}}_i^\dag \mathbf{\Sigma}_{\mathbf{x}} {\mathbf{u}}_i > 0$, for $i=\ell'+1,\ldots,s$ and the \ac{MMSE} is characterized by an error floor in the low-noise regime. Specifically, we can expand the Gaussian \ac{MMSE} as
\begin{equation}
\MMSE^{\sf G}(\sigma^2) =\mathcal{M}^{\sf G}_{\infty} + {\mathcal{D}^{\sf G}} \cdot \sigma^2 + { o}\left(  \sigma^2 \right)
\end{equation}
where
\begin{equation}
 \mathcal{M}^{\sf G}_{\infty} = \sum_{i=\ell' +1}^s  {\mathbf{u}}_i^\dag \mathbf{\Sigma}_{\mathbf{x}} {\mathbf{u}}_i,
\qv \mathcal{D}^{\sf G} = \sum_{i=1}^{\ell'} \frac{1}{\lambda_i} \mathbf{u}_i^{\dag} \mathbf{\Sigma}_{\mathbf{x}} \mathbf{u}_i.
 \end{equation} 

\section{Proof of Theorem~\ref{theo:3}}
\label{app:UpGMM}


The proof is based on the analysis of the upper bound on $\MMSE^{\sf GM} (\sigma^2)$ obtained by considering the mean-squared error associated to the classify and reconstruct decoder described in Section~\ref{par:suff}. The value of such upper bound is given by
\begin{IEEEeqnarray}{rCl}
\nonumber
\MSE_{\sf CR}(\sigma^2) & = &  \sum_k p_k \sum_m p_{\hat{c} | c}(m|k)\\
& & \cdot \E{  \| \mathbf{x} - \mathcal{W}_m (\mathbf{y})  \|^2 | \hat{c}=m,c=k  }\\
 \nonumber
& = & \sum_k p_k p_{\hat{c} | c}(k|k) \E{  \| \mathbf{x} - \mathcal{W}_k ( \mathbf{y}) \|^2 | \hat{c}=c=k  }\\
\nonumber
 & & + \sum_k p_k \sum_{m\neq k} p_{\hat{c} | c}(m|k)  \\
 & & \cdot \E{  \| \mathbf{x} - \mathcal{W}_m(\mathbf{y})  \|^2 | \hat{c}=m,c=k  },
\end{IEEEeqnarray}
where we have denoted by $p_{\hat{c} | c}(m|k)$ the probability that the \ac{MAP} classifier yields $\hat{c}=m$ given that the actual input class  is $c=k$. The first term, then, can be upper bounded by using the law of total probability, and we obtain
\begin{IEEEeqnarray}{rCl}
\nonumber
\MSE_{\sf CR}(\sigma^2) & \leq & \sum_k p_k  \E{  \| \mathbf{x} - \mathcal{W}_k (\mathbf{y})  \|^2 | c=k  }\\
\nonumber
 & & + \sum_k p_k \sum_{m\neq k} p_{\hat{c} | c}(m|k)\\
 &&\cdot  \E{  \| \mathbf{x} - \mathcal{W}_m (\mathbf{y})  \|^2 | \hat{c}=m,c=k  }\\
 \nonumber
 &=& \MSE_{\sf LB}(\sigma^2) \\
 \nonumber
 & &+ \sum_k p_k \sum_{m\neq k} p_{\hat{c} | c}(m|k) \\
 && \cdot \E{  \| \mathbf{x} - \mathcal{W}_m (\mathbf{y}) \|^2 | \hat{c}=m,c=k  },
\end{IEEEeqnarray}
where $\MSE_{\sf LB}(\sigma^2)$ is the lower bound to the \ac{MMSE}, which has been shown to approach zero when $\sigma^2 \to 0$, since we are assuming here that $\ell > s\sub{max}$.

Then, we need to show that, when $m \neq k$,
\begin{equation}
\lim_{\sigma^2 \to 0} p_{\hat{c} | c}(m|k)  \E{  \| \mathbf{x} - \mathcal{W}_m(\mathbf{y})  \|^2 | \hat{c}=m,c=k  } =0,
\label{eq:limint}
\end{equation}
and we consider separately in the remainder of this Appendix the two cases for which the subspaces corresponding to the images of the input covariance matrices of class $k$ and $m$ completely overlap or not. In other terms, we consider separately the cases in which the affine spaces spanned by the signals in class $k$ and class $m$ differ only for a fixed translation or not.


\subsection{Non-overlapping case: $ \frac{s_k + s_m}{2} < s_{km}$}

In this case, given that $\ell > s\sub{max}$, by leveraging the results in~\cite[Theorem 2]{Reboredo13}, we can state that
\begin{equation}
\lim_{\sigma^2 \to 0 }  p_{\hat{c} | c}(m|k) =0.
\label{eq:Perr0}
\end{equation}
The misclassification probability $p_{\hat{c} | c}(m|k)$ is the measure of the set representing the decision region of the \ac{MAP} classifier associated with class $m$ with respect to the Gaussian measure corresponding to the Gaussian distribution
\begin{equation}
\mathcal{N}  \left( 
  \left[
  \begin{array}{cc}
  \boldsymbol{\mu_k} \\
  \mathbf{0}
  \end{array}
  \right], 
\left[
\begin{array}{ccc}
\mathbf{\Sigma}_{\mathbf{x}}^{(k)}   & \mathbf{0}\\
\mathbf{0}   & \sigma^2 \cdot \mathbf{I}
\end{array}
\right]
\right).
\end{equation}
Moreover it can be shown that, in the limit $\sigma^2 \to 0$, the product in (\ref{eq:limint}) is upper bounded by the integral of a measurable function over a set with measure zero, which is then equal to zero.

\subsection{Overlapping case: $ \frac{s_k + s_m}{2} = s_{km}$}


Observe that, in this case, Lemma~\ref{lem:overlap} states that $\mathrm{Im}(\mathbf{\Sigma}_{\mathbf{x}}^{(k)} + \mathbf{\Sigma}_{\mathbf{x}}^{(m)}) = \mathrm{Im}(\mathbf{\Sigma}_{\mathbf{x}}^{(k)}) = \mathrm{Im}( \mathbf{\Sigma}_{\mathbf{x}}^{(m)})$. Then, we further consider two separate cases: 
i) the difference of the mean vectors of classes $k$ and $m$ lies on the subspace spanned by the covariance matrices of the two classes and ii) the difference of the mean vectors of classes $k$ and $m$ does not lie on the subspace spanned by the covariance matrices of the two classes. We consider first the case in which
\begin{equation}
\boldsymbol{\mu}_{\mathbf{x}}^{(k)} - \boldsymbol{\mu}_{\mathbf{x}}^{(m)} \notin \mathrm{Im}(\mathbf{\Sigma}_{\mathbf{x}}^{(k)} + \mathbf{\Sigma}_{\mathbf{x}}^{(m)}).
\end{equation}
For simplicity of notation, we introduce the symbol $\mathbf{M}_{km}=(\boldsymbol{\mu}_{\mathbf{x}}^{(k)} - \boldsymbol{\mu}_{\mathbf{x}}^{(m)}) (\boldsymbol{\mu}_{\mathbf{x}}^{(k)} - \boldsymbol{\mu}_{\mathbf{x}}^{(m)})^\dag$ and we can show that, with probability 1, 
\begin{equation}
\rank ( \mathbf{\Phi}  (  \mathbf{\Sigma}_{\mathbf{x}}^{(k)} + \mathbf{\Sigma}_{\mathbf{x}}^{(m)}    + \mathbf{M}_{km}) \mathbf{\Phi}^\dag ) = s_{km} +1,
\end{equation}
as, by Lemma~\ref{lem:rankim}, $\rank (  \mathbf{\Sigma}_{\mathbf{x}}^{(k)} + \mathbf{\Sigma}_{\mathbf{x}}^{(m)}    + \mathbf{M}_{km} ) = s_{km} +1$ and $\ell > s_{km}$. This implies that, using again Lemma~\ref{lem:rankim}, we have
\begin{equation}
\mathbf{\Phi} (\boldsymbol{\mu}_{\mathbf{x}}^{(k)} - \boldsymbol{\mu}_{\mathbf{x}}^{(m)}) \notin \mathrm{Im}(\mathbf{\Phi}(\mathbf{\Sigma}_{\mathbf{x}}^{(k)} + \mathbf{\Sigma}_{\mathbf{x}}^{(m)})\mathbf{\Phi}^\dag).
\end{equation}
Therefore, by using the result in \cite[Theorem 3]{Reboredo13}, we can state that $\lim_{\sigma^2 \to 0} p_{\hat{c} | c}(m|k) = 0$, and, by using a similar proof to that presented in the previous paragraphs, we can show that (\ref{eq:limint}) is satisfied also in this case.

Consider now the case for which
\begin{equation}
\boldsymbol{\mu}_{\mathbf{x}}^{(k)} - \boldsymbol{\mu}_{\mathbf{x}}^{(m)} \in \mathrm{Im}(\mathbf{\Sigma}_{\mathbf{x}}^{(k)} + \mathbf{\Sigma}_{\mathbf{x}}^{(m)}) = \mathrm{Im}(\mathbf{\Sigma}_{\mathbf{x}}^{(m)}).
\end{equation}
Here, the misclassification probability is not guaranteed to approach zero when $\sigma^2 \to 0$, and we have to resort to a different proof technique. The rationale behind this proof is that also the mismatched mean-squared error approaches zero in the low-noise regime, provided that the signals in the actual class $k$ and the mismatched class $m$ span the same space. 

By using the law of total probability  we can write
\begin{equation}
p_{\hat{c} | c}(m|k)  \E{  \| \mathbf{x} -\mathcal{W}_m (\mathbf{y}) \|^2 | \hat{c}=m,c=k  } \leq \MSE^{\sf MIS}_{km}(\sigma^2)
\end{equation}
where $\MSE^{\sf MIS}_{km}(\sigma^2) = \E{  \| \mathbf{x} -\mathcal{W}_m (\mathbf{y})  \|^2 | c=k  } $ is the mismatched mean-squared error incurred when estimating signals drawn from the Gaussian class $k$ with the conditional mean estimator associated with class $m$.

Observe that, on denoting by $\mathbf{\Sigma}_{\mathbf{y}}^{(k)} = \mathbf{I}\sigma^2 + \mathbf{\Phi} \mathbf{\Sigma}_{\mathbf{x}}^{(k)} \mathbf{\Phi}^\dag$ the covariance matrix of the measurement vector $\mathbf{y}$ conditioned on class $k$, we can write
\begin{IEEEeqnarray}{rCl}
\nonumber
\MSE^{\sf MIS}_{km}(\sigma^2) & = & \E{  \tr \left(    \left( \mathbf{x} -\mathcal{W}_m (\mathbf{y})  \right)   \left( \mathbf{x} -\mathcal{W}_m (\mathbf{y})   \right)^\dag      \right) | c=k  }  \\
\nonumber
& = & \tr \left(  \mathbf{\Sigma}_{\mathbf{x}}^{(k)} \right)  - 2 \tr \left(   \mathbf{W}_k  \mathbf{\Sigma}_{\mathbf{y}}^{(k)} \mathbf{W}_m^\dag \right)   \\
\nonumber
& & \tr \left( \mathbf{W}_m \mathbf{\Sigma}_{\mathbf{y}}^{(k)}  \mathbf{W}_m^\dag    \right)+ \tr \left(  \mathbf{M}_{km} \right)\\ 
\nonumber
& & - 2 \tr \left( \mathbf{M}_{km} \mathbf{\Phi}^\dag \mathbf{W}_m^\dag \right)\\
& & + \tr \left( \mathbf{\Phi} \mathbf{W}_m \mathbf{M}_{km} \mathbf{\Phi}^\dag \mathbf{W}_m^\dag \right).
\end{IEEEeqnarray}
In order to prove that $\MSE^{\sf MIS}_{km}(\sigma^2) $ approaches zero when $\sigma^2 \to 0$, we can show the following four identities:
\begin{equation}
 \lim_{\sigma^2 \to 0}  \tr \left(   \mathbf{W}_k  \mathbf{\Sigma}_{\mathbf{y}}^{(k)} \mathbf{W}_m^\dag \right) =  \tr(\mathbf{\Sigma}_{\mathbf{x}}^{(k)});
 \label{eq:fact1}
 \end{equation}
\begin{equation} 
 \label{eq:fact2}
\lim_{\sigma^2 \to 0}  \tr \left( \mathbf{W}_m \mathbf{\Sigma}_{\mathbf{y}}^{(k)}  \mathbf{W}_m^\dag \right)
 =  \tr(\mathbf{\Sigma}_{\mathbf{x}}^{(k)});
 \end{equation}
 \begin{equation}
  \lim_{\sigma^2 \to 0}  \tr \left( \mathbf{M}_{km} \mathbf{W}_m^\dag \right)
 =  \tr \left(\mathbf{M}_{km} \right);
 \label{eq:fact3}
 \end{equation}
 \begin{equation}
  \lim_{\sigma^2 \to 0}  \tr \left(  \mathbf{W}_m\mathbf{\Phi} \mathbf{M}_{km} \mathbf{\Phi}^\dag \mathbf{W}_m^\dag \right)
 =  \tr \left(\mathbf{M}_{km} \right).
 \label{eq:fact4}
 \end{equation}
 
Specifically, we can leverage the inversion Lemma~\cite{Horn}
\begin{equation}
\mathbf{A} (\mathbf{I}c^{-1} + \mathbf{B}\mathbf{A} )^{-1} \mathbf{B} = \mathbf{I} -( \mathbf{I}  + c \mathbf{A}\mathbf{B})^{-1},
\end{equation}
with $\mathbf{A}=\mathbf{\Phi}^\dag$ and $\mathbf{B}=\mathbf{\Phi} \mathbf{\Sigma}_{\mathbf{x}}^{(m)}$,
%
and write
\begin{IEEEeqnarray}{rCl}
\nonumber
\tr \left(   \mathbf{W}_k  \mathbf{\Sigma}_{\mathbf{y}}^{(k)} \mathbf{W}_m^\dag \right)  & = & \tr \left( \mathbf{\Sigma}_{\mathbf{x}}^{(k)} \mathbf{\Phi}^\dag ( \mathbf{I} \sigma^2 +  \mathbf{\Phi} \mathbf{\Sigma}_{\mathbf{x}}^{(m)}\mathbf{\Phi}^\dag)^{-1}\right. \\
& & \left.  \mathbf{\Phi}   \mathbf{\Sigma}_{\mathbf{x}}^{(m)}  \right)\\
\nonumber
& = & \tr \left( \mathbf{\Sigma}_{\mathbf{x}}^{(k)}  \right)\\
&& - \tr \left( \mathbf{\Sigma}_{\mathbf{x}}^{(k)} ( \mathbf{I} +\frac{1}{ \sigma^2} \mathbf{\Phi}^\dag \mathbf{\Phi} \mathbf{\Sigma}_{\mathbf{x}}^{(m)} )^{-1}    \right). \IEEEeqnarraynumspace
\end{IEEEeqnarray}
Then, by noting that the matrix $\mathbf{\Phi}^\dag \mathbf{\Phi} \mathbf{\Sigma}_{\mathbf{x}}^{(m)}$ is diagonalizable with probability 1, and by following steps similar to those adopted in the proof of Theorem~\ref{theo:Gauss}, we are able to prove (\ref{eq:fact1}). Finally, also (\ref{eq:fact2}), (\ref{eq:fact3}) and (\ref{eq:fact4}) are proved by following a completely similar approach.

\bibliographystyle{IEEEtran}
\bibliography{references_rec}

\begin{IEEEbiography}
[{\includegraphics[height=1.25in,width=1in,clip,keepaspectratio]{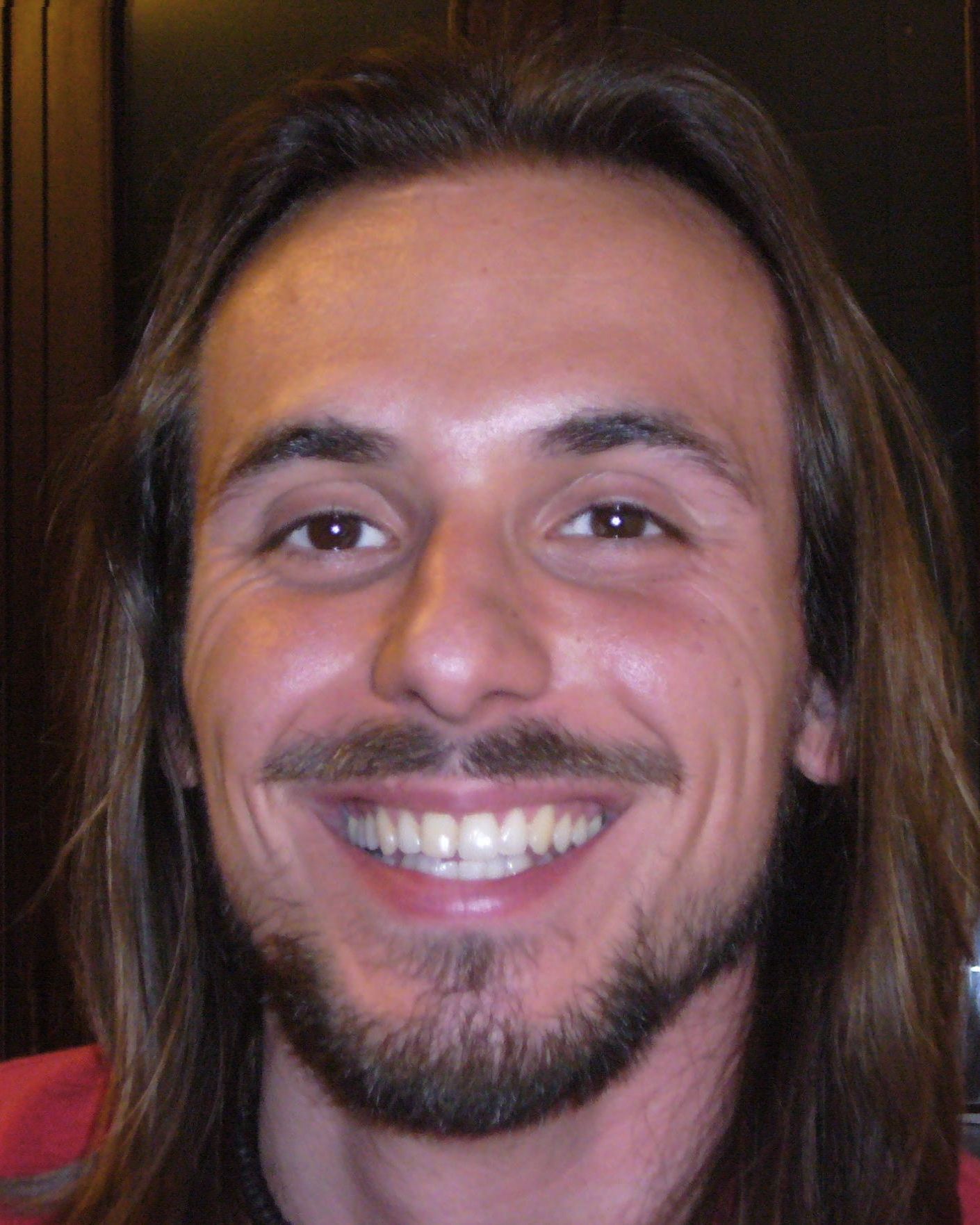}}]
{Francesco Renna} (S'09, M'11) received his Laurea Specialistica Degree in Telecommunication Engineering and Ph.D. degree in Information Engineering, both from University of Padova, in 2006 and 2011, respectively. In 2007 he was an intern at Infineon Technology AG, in Villach, Austria. From 2009 to 2014 he held visiting research appointments with the Princeton University, the Georgia Institute of Technology (Lorraine campus), Supelec, the Duke University and the University College London. In 2011 he was at the Department of Information Engineering of University of Padova as a Research Fellow. Since 2012 he is at the Instituto de Telecomunica\c{c}\~oes, University of Porto, as a Research Fellow. His research interests focus on compressive sensing, physical layer security, multicarrier transmissions and quantum key distribution.
\end{IEEEbiography}
\begin{IEEEbiography}
[{\includegraphics[width=1in,height=1.25in,clip,keepaspectratio]{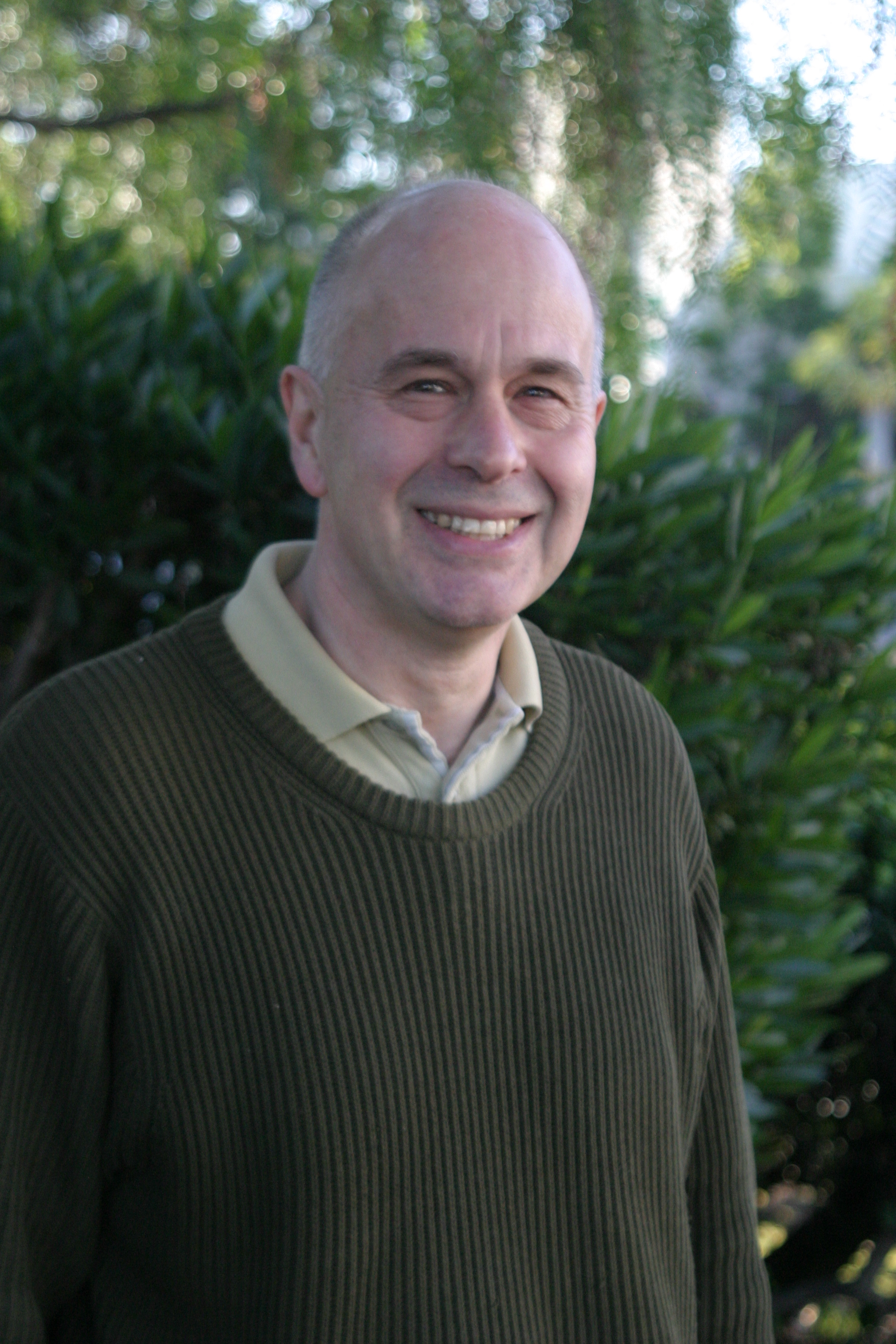}}]
{Robert Calderbank} (M'89, SM'97, F'98) received the BSc degree in 1975 from Warwick University, England, the MSc degree in 1976 from Oxford University, England, and the PhD degree in 1980 from the California Institute of Technology, all in mathematics.
\par
Dr. Calderbank is Professor of Electrical Engineering at Duke University where he now directs the Information Initiative at Duke (iiD) after serving as Dean of Natural Sciences (2010-2013). Dr. Calderbank was previously Professor of Electrical Engineering and Mathematics at Princeton University where he directed the Program in Applied and Computational Mathematics. Prior to joining Princeton in 2004, he was Vice President for Research at AT\&T, responsible for directing the first industrial research lab in the world where the primary focus is data at scale.  At the start of his career at Bell Labs, innovations by Dr. Calderbank were incorporated in a progression of voiceband modem standards that moved communications practice close to the Shannon limit. Together with Peter Shor and colleagues at AT\&T Labs he showed that good quantum error correcting codes exist and developed the group theoretic framework for quantum error correction. He is a co-inventor of space-time codes for wireless communication, where correlation of signals across different transmit antennas is the key to reliable transmission.
 \par
Dr. Calderbank served as Editor in Chief of the IEEE TRANSACTIONS ON INFORMATION THEORY from 1995 to 1998, and as Associate Editor for Coding Techniques from 1986 to 1989. He was a member of the Board of Governors of the IEEE Information Theory Society from 1991 to 1996 and from 2006 to 2008. Dr. Calderbank was honored by the IEEE Information Theory Prize Paper Award in 1995 for his work on the Z4 linearity of Kerdock and Preparata Codes (joint with A.R. Hammons Jr., P.V. Kumar, N.J.A. Sloane, and P. Sole), and again in 1999 for the invention of space-time codes (joint with V. Tarokh and N. Seshadri). He has received the 2006 IEEE Donald G. Fink Prize Paper Award, the IEEE Millennium Medal, the 2013 IEEE Richard W. Hamming Medal, and he was elected to the US National Academy of Engineering in 2005.
\end{IEEEbiography}
\begin{IEEEbiography}
[{\includegraphics[width=1in,height=1.25in,clip,keepaspectratio]{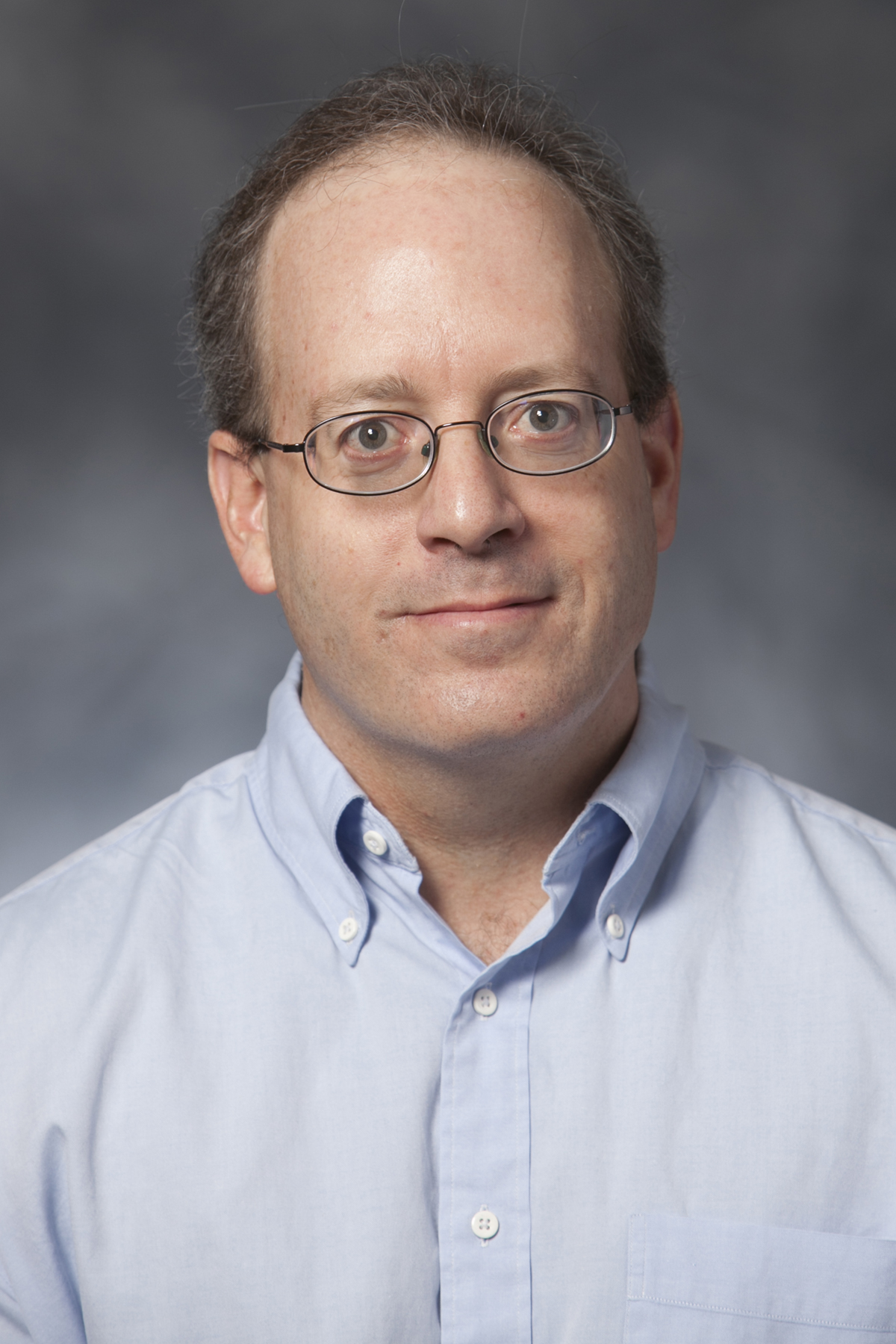}}]
{Lawrence Carin} (SM'96, F'01) earned the BS, MS, and PhD degrees in electrical engineering from the University of Maryland, College Park, in 1985, 1986, and 1989, respectively. In 1989 he
joined the Electrical Engineering Department at Polytechnic University (Brooklyn) as an Assistant Professor, and became an Associate Professor there in 1994. In September 1995 he joined the Electrical Engineering Department at Duke University, where he is now professor and Chairman of the Electrical and Computer Engineering Department. He held the William H. Younger Professorship from 2003-2013 (voluntarily relinquished). He is a co-founder of Signal Innovations Group, Inc. (SIG), a small business. His current research interests include signal processing, sensing, and machine learning. He has published over 250 peer-reviewed papers, and he is a member of the Tau Beta Pi and Eta Kappa Nu honor societies.
\end{IEEEbiography}
\begin{IEEEbiography}
[{\includegraphics[width=1in,height=1.25in,clip,keepaspectratio]{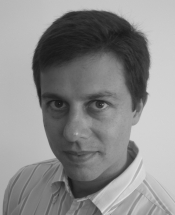}}]
{Miguel R. D. Rodrigues} (S'98, A'02, M'03) received the Licenciatura degree in Electrical Engineering from the University of Porto, Portugal in 1998 and the Ph.D. degree in Electronic and Electrical Engineering from University College London, U.K. in 2002.
\par
He is currently a Senior Lecturer with the Department of Electronic and Electrical Engineering, University College London, UK. He was previously with the Department of Computer Science, University of Porto, Portugal, rising through the ranks from Assistant to Associate Professor, where he also led the Information Theory and Communications Research Group at Instituto de Telecomunica\c{c}\~oes - Porto. He has held postdoctoral or visiting appointments with various Institutions worldwide including University College London, Cambridge University, Princeton University, and Duke University in the period 2003-2013.
\par
His research work, which lies in the general areas of information theory, communications and signal processing, has led to over 100 papers in journals and conferences to date. Dr. Rodrigues was also honored by the IEEE Communications and Information Theory Societies Joint Paper Award in 2011 for his work on Wireless Information-Theoretic Security (joint with M. Bloch, J. Barros and S. M. McLaughlin).
\end{IEEEbiography}

\end{document}